%% file: manuscript.tex
\PassOptionsToPackage{sort&compress}{natbib} % reference ranges

\documentclass[preprint,review,10pt,3p]{elsarticle}

\usepackage{amssymb}
\usepackage{amsmath}
\usepackage{lineno}

\usepackage{xcolor}
\usepackage{lipsum}
\usepackage{hyperref}
\usepackage{tabularx}
\usepackage{booktabs}
\usepackage{multicol}
\usepackage{multirow}
\usepackage{makecell}
\usepackage{subfig}
\usepackage{cleveref}
\usepackage{siunitx}
\usepackage[bottom]{footmisc} % footnotes at bottom of page
\newcommand{\gray}[1]{\textcolor{gray}{#1}}
\usepackage{setspace}
\usepackage{amsmath,amssymb,amsfonts} % for subsequations
\usepackage{relsize} % for mathlarger
\usepackage{gensymb}
\usepackage{commath}
\usepackage{mathrsfs}
\usepackage{bbm}
\usepackage{pifont}
\usepackage{array}

\makeatletter
\newcommand{\vast}{\bBigg@{3}}
\newcommand{\Vast}{\bBigg@{4}}
\makeatother

\newcommand{\customfootnotetext}[2]{{% Group to localize change to footnote
  \renewcommand{\thefootnote}{#1}% Update footnote counter representation
  \footnotetext[0]{#2}}}% Print footnote text

\journal{Applied Energy}

\begin{document}

% Fudge cover page for arXiv
\section*{Quantifying the benefit of load uncertainty reduction for the design of district energy systems under grid constraints using the Value of Information}

\noindent{\large Max Langtry, Ruchi Choudhary}

\begin{figure*}[h]
    \centering
    \includegraphics[width=\linewidth]{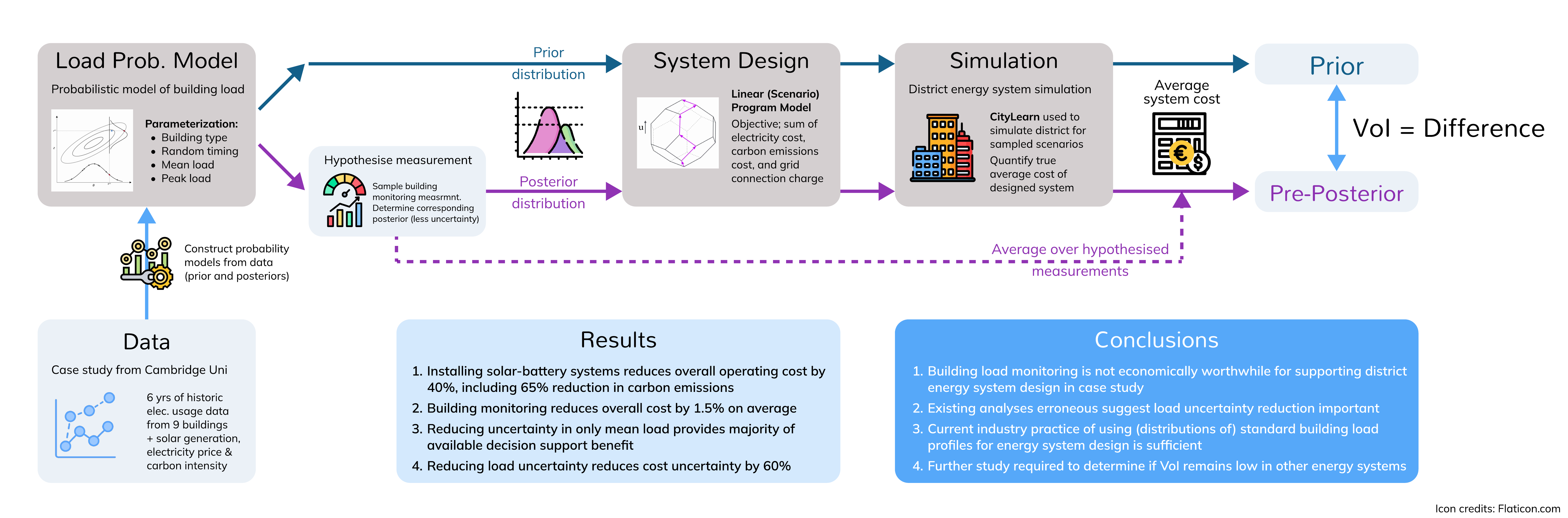}
\end{figure*}

\begin{itemize}
\item Sizing of solar-battery and grid capacity in case study district with historic data
\item Benefit of load uncertainty reduction for supporting energy system design quantified
\item Building monitoring reduces total system costs by less than 1.5\% on average
\item Monitoring found to be not economically worthwhile for supporting system design
\item Reducing only mean load uncertainty provides majority of decision support benefit
\end{itemize}

\newpage

\begin{frontmatter}

%% Title, authors and addresses

%% use the tnoteref command within \title for footnotes;
%% use the tnotetext command for theassociated footnote;
%% use the fnref command within \author or \affiliation for footnotes;
%% use the fntext command for theassociated footnote;
%% use the corref command within \author for corresponding author footnotes;
%% use the cortext command for theassociated footnote;
%% use the ead command for the email address,
%% and the form \ead[url] for the home page:
%% \title{Title\tnoteref{label1}}
%% \tnotetext[label1]{}
%% \author{Name\corref{cor1}\fnref{label2}}
%% \ead{email address}
%% \ead[url]{home page}
%% \fntext[label2]{}
%% \cortext[cor1]{}
%% \affiliation{organization={},
%%             addressline={},
%%             city={},
%%             postcode={},
%%             state={},
%%             country={}}
%% \fntext[label3]{}

\title{Quantifying the benefit of load uncertainty reduction for the design of district energy systems under grid constraints using the Value of Information}
% TBC, try to think of something better
%% Benefit of load uncertainty reduction for design of district energy systems with distributed generation and storage

%% use optional labels to link authors explicitly to addresses:
%% \author[label1,label2]{}
%% \affiliation[label1]{organization={},
%%             addressline={},
%%             city={},
%%             postcode={},
%%             state={},
%%             country={}}
%%
%% \affiliation[label2]{organization={},
%%             addressline={},
%%             city={},
%%             postcode={},
%%             state={},
%%             country={}}

\author[EECi]{Max Langtry\corref{cor1}} \ead{mal84@cam.ac.uk}
\author[EECi,ATI]{Ruchi Choudhary}

\cortext[cor1]{Corresponding author}

%% Author affiliation
\affiliation[EECi]{organization={Energy Efficient Cities Initiative, Department of Engineering, University of Cambridge},
            addressline={Trumpington Street},
            city={Cambridge},
            postcode={CB2 1PZ},
            country={UK}}
\affiliation[ATI]{organization={Data-Centric Engineering, The Alan Turing Institute},
            addressline={British Library},
            city={London},
            postcode={NW1 2DB},
            country={UK}}

%% Abstract
\begin{abstract}
%% Text of abstract
% 250 words
Load uncertainty must be accounted for during design to ensure building energy systems can meet energy demands during operation. Reducing building load uncertainty allows for improved designs with less compromise to be identified, reducing the cost of decarbonizing energy usage. However, the building monitoring required to reduce load uncertainty is costly.

This study uses Value of Information analysis (VoI) to quantify the economic benefit of practical building monitoring for supporting energy system design decisions, and determine if its benefits outweigh its cost.
An extension of the VoI framework, termed `On-Policy' VoI, is proposed, which admits complex decision making tasks where decision policies are required. This is applied to a case study district energy system design problem, where a Linear Program model is used to size solar-battery systems and grid connection capacity under uncertain building loads, modelled using historic electricity metering data.

Load uncertainty is found to significantly impact both system operating costs ($\pm30\%$) and the optimal system design ($\pm20\%$). However, using building monitoring data to improve the design of the district reduces overall costs by less than $1.5\%$ on average. As this is less than the cost of measurement, using monitoring is not economically worthwhile in this case. This provides the first numerical evidence to support the sufficiency of using standard building load profiles for energy system design.
Further, reducing only uncertainty in mean load is found to provide most of the available decision support benefit, meaning using hourly measurement data provides little benefit for energy retrofit design.
\end{abstract}

%% Keywords
\begin{keyword}
%% keywords here, in the form: keyword \sep keyword
District energy system \sep Uncertainty reduction \sep Value of Information \sep Energy planning \sep Stochastic programming \sep Building monitoring
\end{keyword}

\end{frontmatter}

%% Add \usepackage{lineno} before \begin{document} and uncomment 
%% following line to enable line numbers
%\linenumbers

%% main text - aim for 20 pages

\input{sections/intro}
\input{sections/theory}
\newpage
\input{sections/experiment}
\newpage
\input{sections/results}
\newpage
\input{sections/conclusions}
\newpage
\input{sections/endmatter}

%% The Appendices part is started with the command \appendix;
%% appendix sections are then done as normal sections
\newpage
\appendix
\input{sections/appendices}

%% If you have bib database file and want bibtex to generate the
%% bibitems, please use
%%
\newpage
\bibliographystyle{elsarticle-num} 
\bibliography{BD-VOI_refs}

\end{document}

%% file: sections/intro.tex
\section{Introduction} \label{sec:intro}

% aim for c. 1500 words

\subsection{Background}

% Outline problem of heat electrification in building energy systems increasing load peaking which is issue given grid capacity constraints - quickly discuss context of grid constraints arising as decarbonisation progresses
% Discuss solar-battery systems are route to decarbonisation while managing grid usage - local decarbonisation without big grid impact (battery needed for peak shaving)
% Discuss issue of uncertainties affecting future buidling energy systems
% Mention that some of these uncertainties can be reduced through measurement, specifically interested in load profiles in this case, as demand peaks caused be heating could be a significant factor for grid impact

Building energy systems accounted for over 40\% of final energy consumption and 35\% of carbon emissions in the EU in 2020 \cite{eea2023DecarbonisingHeatingCooling}.
Hence, decarbonizing building energy usage is essential for achieving targets of net-zero carbon emissions by 2050 \cite{committeeonclimatechange2020SixthCarbonBudgetb}.
As space and water heating makes up around 75\% of final energy usage in buildings, it is necessary to install low-carbon heating systems to replace existing gas-based systems.
Many countries such as the UK and EU have implemented policies that focus on the adoption of heat pumps to decarbonize building heating through electrification \cite{houseofcommons2022DecarbonisingHeatHomes,iea2021NetZero2050}.
% There is also concern about the cost of decarbonising electricity \cite{beis2023PoweringBritainNet}
% Heat electrification increases load and is a challenge for the grid
% Grid connection significant concern for decarbonisation, potential bottleneck ...
However, facilitating widespread roll-out of heat pumps poses significant challenges for the electrical grid. Electrifying heating demand will substantially increase total electricity demand, and more significantly peak electricity demand \cite{heinen2018HeatElectrificationLatest}, which would rise by 50\% for the case of the UK \cite{zhang2022AssessmentImpactsHeat,charitopoulos2023Impact100Electrification}. Supporting this additional electricity demand while maintaining security of supply requires investment in distribution and transmission systems \cite{blonsky2019PotentialImpactsTransportation}, as well as reserve generation capacity \cite{charitopoulos2023Impact100Electrification}, imposing additional costs to heat decarbonisation. Already, capacity limitations on the electrical supply system are causing delays to both building construction and renewable power generation projects \cite{beis2023PoweringBritain,departmentforbusinessandtrade2023UKBatteryStrategy}, and it is widely acknowledged that network capacity limitations pose a significant barrier to future decarbonisation \cite{haben2023ADViCEAIDecarbonisation}.

% So, decarbonized building energy systems must be designed to minimise grid impact
% Distributed generation and storage is key way of achieving this
To achieve decarbonization by 2050, energy retrofits must be cheap enough so that they are widely adopted, but also able to be safely integrated into the wider energy system. Hence, building energy systems must be designed to balance their total system cost, the carbon emissions reductions they achieve, and their impact on the electrical grid.
The use of distributed generation and storage has emerged as a prominent pathway for reducing the carbon emissions of building energy usage, and so has attracted considerable research interest in recent literature \cite{li2023ReviewPhotovoltaicBattery,niveditha2022OptimalSizingHybrid,sharma2020EconomicPerformanceAssessment,novoa2019OptimalRenewableGeneration,salvador2012MethodologyDesignEnergy,tumminia2020GridInteractionEnvironmental}.
This is due to the falling costs of solar PV generation \cite{irena2023RenewablePowerGeneration}, and the need for local storage to both improve buildings' utilization of the solar energy and reduce the grid impact caused by peaks in solar generation \cite{li2023ReviewPhotovoltaicBattery}.

% Discuss the widespread acknowledgement of the importance of uncertainties for DES design (lots of uncertainties affect DESs, widely acknowledged, etc., etc.) (also mention complex uncertainties, e.g. funcational variables) - mention review papers
When designing building energy systems, there is substantial uncertainty concerning the conditions in which they will operate. These include weather conditions which affect local energy generation, building usage type and occupant behaviour which determines the timing of energy usage, and building envelope characteristics which influence the building thermal response. Of particular importance for the management of grid impacts is the electricity usage behaviour of the building, including the timing and magnitude of peak loads. Due to the greater weather dependence and seasonal variability of heating loads, electrification of heating will increase uncertainty in building electricity usage \cite{deroubaix2021LargeUncertaintiesTrends,egging-bratseth2021SeasonalStorageDemand}, in terms of both the aggregate and peak demand.
The challenges posed by these uncertainties, and methods for properly accounting for them during the design of building energy systems have been widely studied \cite{zhu2022UncertaintySensitivityAnalysis,decarolis2017FormalizingBestPractice,fodstad2022NextFrontiersEnergy,mavromatidis2018ReviewUncertaintyCharacterisation,tian2018ReviewUncertaintyAnalysis}.

% Two options for better desgins: better methods or reduced uncertainty
% Reducing uncertainties e.g. through modelling or measurement enables better design, but is costly (justify costs)
% But no studies look at the benefits of this for building energy systems, or whether it is economically worthwhile
Two routes exist for improving the design, and so eventual operational performance, of future decarbonized building energy systems. Firstly, improved optimization methods for designing energy systems in the presence of uncertainties can be developed. Secondly, the uncertainty in the conditions in which the energy system will operate can be reduced by monitoring the heating and electricity use patterns of the building prior to designing the heating retrofit. Optimization methodologies have received concerted research effort for many years \cite{fodstad2022NextFrontiersEnergy,connolly2010ReviewComputerTools}, whereas there has not yet been any study of the potential benefits of uncertainty reduction to improve energy system design.

Reducing uncertainty is advantageous to design as it reduces the need to adopt compromise designs that provide good performance across a wider range of operational scenarios, which causes worsened performance under the conditions the system eventually faces. The more accurate the statistical understanding of the operational conditions, the more tailored a system can be, and the lower its cost.
However, uncertainty reduction is itself costly. For the case of existing buildings, these costs arise from the hardware and software needed to monitor energy consumption patterns, and the cost of management, curation, analysis, and interpretation of the data.
In the UK, only 51\% of electricity and gas meters are currently `smart' (able to collect hourly energy usage data), despite rollout beginning in 2012, and expected to cost £19.4bn in total \cite{desnz2023UpdateRolloutSmart}.
Therefore, it becomes important to quantify the benefits of uncertainty reduction through monitoring buildings on energy system design. Specifically, whether the reductions in total system cost are greater than the cost of gathering and exploiting the data needed to reduce uncertainty.

\subsection{Quantifying the impacts of uncertainty on building energy system design}

% Discuss literature related to quantifying benefit of uncertainty reduction from measurement - highlight this is specific focus, but most literature only looks at initial uncertainties, not reduction
% Importantly, within discussion of limitations of existing methods, discuss how most methods look at the impact of uncertainty on the objective (useful for quantify risk), but VoI looks at the impact of uncertainty on regret, which is much more pertinent for decision making - it matters whether our decision would change, and how much better we could do! (interpretation is important, make this really clear)

% No literature directly considering benefit of uncertainty reduction on design process for energy systems
% As a results, none consider the question of whether expenditure on uncertainty reduction is worthwhile for the design process
% However, this has been studied in other fields - see Lit Review notes
Study of the effects of uncertainty reduction on the design of building energy systems, and the quantification its benefits for decision making, have only recently been proposed \cite{langtry2024RationalisingDataCollection,niu2023FrameworkQuantifyingValuea}.
However, the opportunities for uncertainty reduction to improve the design of engineering systems have been explored in other fields. For example, \cite{acar2009SystemReliabilityBased} quantifies the improvements in the reliability and weight of vehicle designs optimized for crash worthiness when reducing uncertainty in structural material properties, crash performance estimates (from simulations), and experimental tolerances. It proposes that the quantified trade-off can be used to ``decide whether to allocate \ldots resources for employing uncertainty reduction measures''.
% Discuss resource allocation (experiment (reduce uncertainty) vs simulate (exploit) papers \cite{jiang2016ReductionEpistemicModel,ungredda2022BayesianOptimisationVs}??

% Loads of techniques looking at sensitivity of solutions & models to uncertainties (i.e. how do solutions perform under uncertainties), and how to make solutions that are robust to uncertainties - discuss these methods and how they handle/investigate the impact of uncertainties (see MRes project Lit Review for methods)
% Keep this brief and discuss big idea that these studies only look at impact of uncertainties on performance (objective), and not the optimal design
Within the building energy systems literature, the impact of uncertainties on the performance of energy systems has been widely studied.
Uncertainty Analysis \cite{mavromatidis2018ReviewUncertaintyCharacterisation,tian2018ReviewUncertaintyAnalysis,prataviera2022EvaluationImpactInput,gang2015ImpactsCoolingLoad} and Sensitivity Analysis \cite{mavromatidis2018ReviewUncertaintyCharacterisation,prataviera2022EvaluationImpactInput,mavromatidis2018UncertaintyGlobalSensitivity,liu2021FrameworkUncertaintySensitivity} have been applied to analyse the uncertainty in the performance of energy systems caused by uncertainty in operating conditions and building models. For example, \cite{prataviera2022EvaluationImpactInput} uses Sensitivity Analysis to identify the temperature set point and building wall area as the key uncertain parameters affecting estimated energy usage for heating.
However most commonly the impacts of uncertainty are considered in the context of system design \cite{khezri2022OptimalPlanningSolar,oshaughnessy2018SolarOptimizationDistributed,sadeghibakhtiar2024SizeOptimizationStandalone,pickering2019DistrictEnergySystem}. A diverse range of methods for accounting for uncertainties during energy system design optimization have been proposed, and several reviews have been conducted to classify and compare the different approaches \cite{zhu2023ReviewDistributedEnergya,liu2020EnergySystemOptimization,alonso-travesset2022OptimizationModelsUncertainty}. Broadly, these methods provide a representation of the uncertainty in the system performance (design objective) and/or design space (feasible region) caused by the underlying uncertainties in the energy system, and seek to identify a single system design (optimal solution) that either performs best on average or has a low risk in its performance. Whilst these methods account for the influence of uncertainty on candidate designs, they are not able to quantify its impact on the ex-post optimal design.

% A few papers in some sense look at the impact of uncertainty on the optimal system design
% - mavromatidis2018UncertaintyGlobalSensitivity; performs Uncertainty Analysis (UA) (probabilistic analysis) and finds the optimal design on a DES for each sample, uses this to take a broad view on the dominant technology choices, but to some extent shows impact of uncertainties on design process
% - sun2015SensitivityAnalysisMacroparameters; performs (deterministic) UA (local sensitivity analysis) looking at how the optimal size of each component in a BES changes as the value of uncertain parameters change (see paper for parameters)
% - sanajaobasingh2018ModelingSizeOptimization; does similar to above for a different DES
% These papers show the start of the idea, i.e. how do uncertainties affect the optimal design, but they are very limited and do not progress to considering the impact on regret (briefly defined regret and justify importance for decision making)
% To some extent MGA (near-optimal feasible regions) also kind of fits in here, but that's less about modelling uncertain parameters and more about exploring the design space for robustness - need to think how to discuss this
Study of the effects of uncertainty on design has been very limited in the energy systems literature, with only a few previous works investigating the influence of uncertain parameters on the optimal system design.
\cite{sun2015SensitivityAnalysisMacroparameters} analyses the design of a Net Zero Energy Building with distributed generation and storage, and plots how the capacities of each system component (chillers, wind turbines, solar panels, and battery) and the total cost of the optimized system change with uncertain values of properties of the building (including wall thickness, internal heat gain, temperature set point). Additionally, influence coefficients of the uncertain parameters are computed for the capacity of each system component. A later study, \cite{sanajaobasingh2018ModelingSizeOptimization}, performed a similar analysis for another system with distributed generation and storage, investigating how the optimal design and total cost changed with varying scenarios of renewable generation. In both cases no statistical distributions of the uncertain parameters are considered, and so the insights gained are limited. E.g. for the case of \cite{sanajaobasingh2018ModelingSizeOptimization}, it is unclear whether a design with a substantially greater number of PV panels is likely to be optimal, and so worth investing in.

A statistical Sensitivity Analysis for the optimal design of an energy system with distributed generation and storage was performed in \cite{mavromatidis2018UncertaintyGlobalSensitivity}. In this study, a Mixed Integer Linear Program was used to optimize the sizing of boilers, heat pumps, thermal storage, and PV generation for scenarios defined by samples from distributions of uncertain parameter values, including energy and capital costs, conversion efficiencies, and energy demand. This enabled a distribution of optimal sizes to be created for each component to illustrate how the uncertainties affect the optimal design choice. \cite{pickering2019DistrictEnergySystem} performed a similar procedure for the sizing of components in a district energy system, however the optimized designs were not analyzed. Whilst these studies identify the distribution of optimal designs resulting from the underlying uncertainties in the system, they do not then provide a means to quantify the importance of those uncertainties for the design process.

\subsection{Value of information analysis}

% Explain VoI and how it is applicable to addressing this problem - explain how VoI can be used to quantify the impact of uncertainties on the design process
% Discuss existing literature from other fields
% Also mention old school VoI for capacity expansion planning (large-scale energy systems) & limitations (no relation to actual measurements)
Value of Information analysis (VoI) \cite{raiffa1969ReviewDecisionAnalysis,howard1966InformationValueTheory} is a branch of Bayesian Decision Analysis that provides a framework quantifying the expected (mean) benefit to decision making derived from the reduction of epistemic uncertainty (uncertainty associated with lack of knowledge of a system \cite{zhang2021ValueInformationAnalysis}) through measurement. Its general methodology can be used to quantify the benefit of uncertainty reduction for any decision making problem under uncertainty, provided a mathematical description of the decision problem can be produced, i.e. the decision can be formulated as a stochastic decision problem \cite{pratt1995IntroductionStatisticalDecision}.
VoI has been applied in fields such as environmental science, agriculture, and economics \cite{keisler2014ValueInformationAnalysis} to quantify the economic benefit provided by uncertainty reduction interventions to improve decision making.
Within engineering, VoI has been used to determine whether proposed measurements and the uncertainty reduction they provide are financially worthwhile for supporting decisions regarding maintenance scheduling for buildings \cite{grussing2018OptimizedBuildingComponent} and wind farms \cite{myklebust2020ValueInformationAnalysis}, construction project planning \cite{esnaasharyesfahani2020PrioritizingPreprojectPlanning}, and structural health monitoring \cite{difrancesco2021DecisiontheoreticInspectionPlanning,difrancesco2023SystemEffectsIdentifying}.

% Discuss relevance of my paper (only proper VoI in context of building energy systems) - does quantify effect and benefit of uncertainty reduction for system design, but with extremely simple model
% Critique that one building energy system VoI paper \cite{niu2023FrameworkQuantifyingValuea} - does not related to measurement, quantification is rather meaningless (uncertainties not epistemic and cannot be reduced)
Only two previous studies have applied VoI to study the benefit of uncertainty reduction for decision making in building energy systems.

The first, \cite{langtry2024RationalisingDataCollection}, provides a proof-of-concept study of the capabilities of VoI to support decision making across three different building scenarios, quantifying the economic benefit provided by different measurements.
The key limitation of this study is the simplicity of the modelling performed, which does not capture the full complexity of the uncertainties, decision options, and costs.
Additionally, the investigation of uncertainty reduction in each scenario considers only a single uncertain parameter, providing guidance for one measurement intervention in each case.

The second, \cite{niu2023FrameworkQuantifyingValuea}, studies the influence of uncertainties on the design of a district energy system with distributed generation and storage. It quantifies the expected reduction in the total lifetime cost (capital investment plus operational cost) achieved by eliminating uncertainty in building energy demand and solar generation, to improve the sizing of system components (including chillers, solar panels, and energy storage).
Whilst the mathematical VoI framework and literature are not acknowledged in this paper, the study computes the Expected Value of Perfect Information (EVPI), defined in Section \ref{sec:voi}, for the decision problem of designing a district energy system, using a high temporal resolution (hourly) model.
However, the value of information quantified in this study was not related to any practical measurement intervention to reduce uncertainty. In fact, the uncertainties considered in the analysis are not only epistemic, as at the time of design the energy demand and solar generation during operation cannot be known perfectly. Therefore, quantifying the benefit of eliminating these uncertainties provides limited practical insight, particularly as the value of removing uncertainty was found to be significant, and so is not informative as an upper bound.
Similarly, a number of older studies investigating large-scale energy systems (transmission to national scale) quantify the benefit of eliminating uncertainty without association to any practical measurement \cite{modiano1987DerivedDemandCapacity,krukanont2007ImplicationsCapacityExpansion,fursch2014OptimizationPowerPlant,wendling2019BridgesRenewableEnergy}.

% Conclude - no investigations of opportuntity for uncertainty reduction via **measurements** to improve building energy system design - and VoI is perfectly suited to study this problem (it is a framework for doing so)
There remains a substantial research gap regarding the quantification of the benefits of uncertainty reduction to support the design of building energy systems, and the identification of opportunities for economically worthwhile measurement interventions to improve design. Value of Information analysis provides a methodology for addressing this gap, and its ability to do so has already been demonstrated \cite{langtry2024RationalisingDataCollection}.

\newpage
\subsection{Research objectives \& novel contributions}

% Usual structure
% Story is about buildings and uncertainty reduction (data collection) to improve design, VoI is just the tool, but mention that On-Policy VoI is novel, if a very minor tweak of the framework

% Research gap is study of uncertainty reduction to improve decision making for design of building energy systems - specific application is load monitoring for design of district energy system under grid constraints

% Novelties are:
% - First quantification of the benefit of load monitoring (or actually any measurement) for DES design
% - On-Policy VoI framework
% - ? First demonstration of existing building data being sufficient for design of new systems
% - ? Quantification of relative importance of shape, mean, peak uncertainties for system design w.r.t. induced regret (pertinent metric for decision making)
% Highlight importance to practitioners - i.e. provides recommendations on economic measurements to support design - key importance is this work relates to an *actual* measurement and associated uncertainty reduction

In the existing literature, there has been very limited study of the economic benefit of uncertainty reduction from measurement (data collection) to support decision making in building energy systems. No previous study has quantified the benefit of using hourly energy metering data to support the design of district energy systems by reducing uncertainty in building load profiles.
Understanding of the value of load monitoring for improving energy system design is necessary to determine whether significant cost savings are currently being missed, or whether current industry practice of using standard load profiles for design is sufficient.

\begin{enumerate}
    \item[] This study applies a novel extension of the Value of Information analysis methodology, termed `On-Policy' VoI, to quantify the overall cost savings achieved when using hourly building monitoring data to inform energy system design. It investigates the sizing of solar-battery systems and grid connection capacity for an illustrative district energy system, where probabilistic models of building load profiles are constructed using historic energy usage data from the Cambridge University estate.
    The main objectives of this study are:
    \begin{itemize}
        \item Determine if load monitoring is economically worthwhile for improving energy system design by reducing load uncertainty
        \item Quantify the relative importance of load profile features (mean load, peak load, and shape) for improving system design
        \item Gather evidence to assess whether current industry practice of using standard load profiles for building energy system design is sufficient
    \end{itemize}
\end{enumerate}

This work is the first to quantify the economic value of practical building measurement to support energy system design through load uncertainty reduction. It demonstrates that existing methods for assessing the impact of uncertainties on energy system design are insufficient, and lead to erroneous conclusions about the importance of uncertainty reduction. Further, it provides the first numerical evidence to indicate that existing industry practice of using (distributions of) standard building load profiles for energy system design is adequate for minimizing the cost of energy.
% Add extra comments about importance of contribution??

The remainder of this work is structured as follows. Section \ref{sec:theory} outlines the Value of Information analysis (VoI) framework, and presents the `On-Policy' extension which enables its application to complex decision making problems. Section \ref{sec:experiment} describes the case study district energy system. Sections \ref{sec:data} \& \ref{sec:prob} present the data studied, and how it is used to construct probabilistic models of building loads. Next, Sections \ref{sec:SP} \& \ref{sec:sims} define how the system is designed using a Linear (Scenario) Programming model, and evaluated via simulation. The VoI calculations are performed in Section \ref{sec:results}, and the practical importance of the results obtained is discussed. Finally conclusions are drawn in Section \ref{sec:conclusions}.

%% file: sections/theory.tex
\section{Methodology} \label{sec:theory}

Value of Information analysis (VoI), initially introduced by Raiffa \cite{raiffa1969ReviewDecisionAnalysis} and Howard \cite{howard1966InformationValueTheory} in the 1960s, is a framework based on Bayesian Decision Analysis and Expected Utility Theory \cite{smith1945TheoryGamesEconomic} for quantifying the expected benefit to decision making provided by the reduction of epistemic uncertainty (uncertainty associated with lack of knowledge of a system \cite{zhang2021ValueInformationAnalysis}) through data collection (measurement). It numerically answers the question, ``How much is decision making improved on average by reducing the epistemic uncertainty in a given problem?''.
This section briefly introduces the existing VoI framework, before then extending it in Section \ref{sec:on-policy} to enable its application to complex decision making problems, including the design of district energy systems.
A thorough explanation of the classical VoI framework is available in \cite{zhang2021ValueInformationAnalysis}.

\subsection{Bayesian Decision Analysis}

Bayesian Decision Analysis provides a mathematical framework for studying decision-making in the presence of uncertainties, particularly circumstances where those uncertainties can be reduced through measurement. Its aim is to determine the optimal set of actions which should be taken by a decision-maker (termed an `actor') in order to maximise their expected utility That is, find the decision which when taken in the system provides the highest reward/benefit to the decision-maker on average over the uncertainties in the problem. This task can be formulated as a mathematical (stochastic) optimization problem.

Consider a generalised stochastic decision problem in which an actor seeks to select a `decision action' to take, $a \in \mathcal{A}$, within a system with uncertain parameters $\theta$, which have a prior probabilistic model (distribution), $\pi(\theta)$. The performance/benefit of each available action is described by a utility function which is also dependent upon the uncertain parameters, $u(a,\theta)$. In VoI analysis, before an action $a$ is taken, the actor may choose to take a `measurement action', $e \in E$, which provides the actor with some data $z$ that reduces uncertainty in $\theta$. The probabilistic model describing the measurement data $f_e(z|\theta)$ is used to update the prior model, $\pi(\theta)$, to produce a posterior probabilistic model (distribution), $\pi(\theta|z)$. This posterior (which has reduced uncertainty c.f. the prior) is then used by the actor to inform their choice of `decision action', improving their decision making performance.

The set of available actions, prior probabilistic model, and utility function, $\lbrace\mathcal{A},\pi(\theta),u(a,\theta)\rbrace$, provide a complete mathematical description of the decision making task under uncertainty. The likelihood function $f_e(z|\theta)$ describes the reduction in epistemic uncertainty in the parameters of the system, $\theta$, provided by data collection.

\begin{figure}[h]
    \centering
    \includegraphics[width=0.8\linewidth]{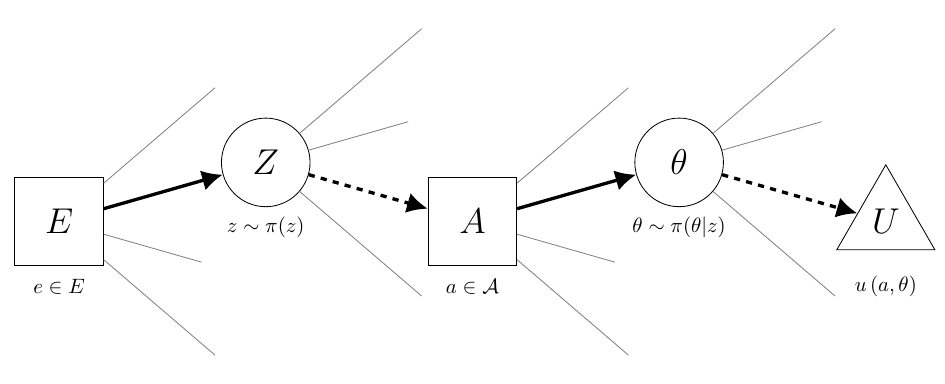}
    \caption{Decision tree representation of Pre-Posterior Decision Problem}
    \label{fig:DT-prepost}
\end{figure}

This generalised model can be represented in decision tree form, as shown in Fig. \ref{fig:DT-prepost}, in which square nodes represent decisions, circular nodes represent uncertainties, and triangular nodes represent utilities.

\newpage
The actor, who is assumed to be risk neutral, aims to maximise the expected utility they receive from the `decision action' $a$ they select. Costs are defined as negative utilities. The actor may choose to do this without taking any measurement. The resulting stochastic optimisation is termed the Prior Decision Problem,
\begin{equation}
    \max_{a \in \mathcal{A}} \, \mathbb{E}_{\theta} \left\lbrace u(a,\theta) \right\rbrace
\end{equation}

Alternatively, the actor can initially take a `measurement action' $e$ which reduces uncertainty in the problem, and improves their subsequent choice of `decision action'. In this case, the optimization of expected utility is performed over both measurement and decision actions, and is termed the Pre-Posterior Decision Problem,
\begin{equation}
    \max_{e \in E} \, \mathbb{E}_{z} \left\lbrace \max_{a \in \mathcal{A}} \, \mathbb{E}_{\theta|z} \left\lbrace u(a,\theta) \right\rbrace \right\rbrace
\end{equation}

\subsection{Value of Information} \label{sec:voi}

By taking a measurement and reducing uncertainty in the problem the actor is able to improve their decision making, i.e. increase the average utility they obtain from the decision problem. This increase in expected utility provided by uncertainty reduction from measurement (the difference in expected utilities achieved when decisions are made with and without uncertainty reduction respectively) is termed the Value of Information (VoI) \citep{raiffa1969ReviewDecisionAnalysis}.

If a measurement $e$ is taken which provides imperfect/uncertain information that reduces but does not remove the epistemic uncertainty in the parameters $\theta$, the expected value of that uncertain information is termed the Expected Value of Imperfect Information (EVII), and is given by,

\begin{equation} \label{eq:EVII}
    \text{EVII}(e) = \mathbb{E}_{z} \left\lbrace \max_{a \in \mathcal{A}} \: \mathbb{E}_{\theta|z} \left\lbrace u(a,\theta) \right\rbrace \right\rbrace - \max_{a \in \mathcal{A}} \: \mathbb{E}_{\theta} \left\lbrace u(a,\theta) \right\rbrace
\end{equation}

where the posterior probabilistic model, $\pi(\theta|z)$, used to compute inner expectation with respect to $\theta|z$, is derived using the likelihood function for the measurement $e$, $f_e(z|\theta)$.

In the case where the measurement provides perfect information on the true value of the uncertain parameters of the system, the actor is able to make the best possible decision, and the expected utility simplifies to,
\begin{equation}
\mathbb{E}_{z} \left\lbrace \max_{a \in \mathcal{A}} \, \mathbb{E}_{\theta|z} \left\lbrace u(a,\theta) \right\rbrace \right\rbrace \: \longrightarrow \: \mathbb{E}_{z} \left\lbrace \max_{a \in \mathcal{A}} \, u(a,z) \right\rbrace
\end{equation}

The expected value of information in this case is termed the Expected Value of Perfect Information (EVPI), and is given by,
\begin{equation} \label{eq:EVPI}
    \begin{aligned}
        \text{EVPI} &= \mathbb{E}_{z} \left\lbrace \max_{a \in \mathcal{A}} \, u(a,z) \right\rbrace - \max_{a \in \mathcal{A}} \mathbb{E}_{\theta} \left\lbrace u(a,\theta) \right\rbrace \\[1em]
    \end{aligned}
\end{equation}
The EVPI is significantly cheaper to compute, and provides an upper-bound on the EVII \cite{keisler2014ValueInformationAnalysis}.\\

The Value of Information quantifies how much better on average the decision-maker is able to do at making their decision by reducing uncertainty in the parameters of the system. Typically, decision problems are formulated using an economic objective, i.e. a total cost or profit. Hence, the VoI quantifies the actor's willingness to pay for measurements that reduce uncertainty in the decision problem. Comparing VoI to the cost of reducing uncertainty allows the net economic benefit to be quantified, i.e. the VoI minus the cost of information. With this, a decision maker can determine whether the measurements considered are economically worthwhile, and compare the relative benefit of different uncertainty reduction options.

% Necessary? My case has remaining uncertainty which perfect measurement, so perhaps
Most practical engineering systems contain aleatoric uncertainty (inherent variability which cannot be explained away) as well as epistemic uncertainty. This aleatoric uncertainty cannot be reduced via measurement, and so even when all uncertainty in $\theta$ is removed from the system, some residual uncertainty remains. To account for this, the parameters $\theta$ are defined as the measurable parameters of the system (which may contain both epistemic and aleatoric uncertainty, meaning it may not be possible to fully remove their uncertainty), and additional unmeasureable parameters $\phi$ are introduced. So the utility function is now,
\begin{equation}
    u(a,\theta) = \mathbb{E}_{\theta,\phi} \left\lbrace u(a,\theta,\phi) \right\rbrace
\end{equation}

\newpage
\subsection{On-Policy Value of Information} \label{sec:on-policy}

% Introduce On-Policy VoI and discuss nuances of interpretation required

The current VoI framework assumes that it is possible to determine the optimal action, and that in solving the stochastic optimisation problem a sufficient approximation of the expected utility is computed.
% ... i.e. the following optimisation is tractable or well approximatable ...
% \begin{equation}
%     \max_{a \in \mathcal{A}} \, \mathbb{E}_{\vartheta \sim \pi(\cdot)} \left\lbrace u(a,\vartheta) \right\rbrace
% \end{equation}
However, many practical engineering decision problems are too complex for direct stochastic optimisation to be tractable. Instead, decisions are determined via policies, such as Rule-Based Control, Reinforcement Learning, or Stochastic Programming. These policies achieve tractability by either using heuristics to guide their search, or by simplifying the model of the energy system and/or the statistics used to evaluate performance\footnote{Most notably, Monte Carlo approximations of the expected utility using only a small number of samples are typically used to limit computational cost, but come at the expense of significant statistical error.}. As a result, policies may not provide a sufficiently accurate (low error) approximation of the decision maker's target utility to capture useful VoI values, i.e. errors are significantly greater than meaningful VoI values.\\

The classical VoI framework is extended to allow decision making via policies. This extension involves an additional step of evaluating the expected utility achieved by actions selected by the policy.

A decision policy is defined as a mapping from a utility function describing the performance of the system, $u(a,\theta)$, and a distribution describing the uncertainty in the system parameters, $\pi(\theta)$, to an action.
\begin{equation}
    \mathcal{P} \left( u(a,\theta),\pi(\theta) \right) \rightarrow \alpha \in \mathcal{A}
\end{equation}

As the utility function and action set are defined by the given stochastic decision problem, the policy applied to the decision problem is referred to as $\mathcal{P}_u(\pi(\theta))$.\\

% Give plain and clear explanation of how decisions are made and evaluations are performed
% Need to evaluate expected performance of policy w.r.t true utility
As before, the VoI is the difference between the expected utilities achieved when decisions are made with and without uncertainty reduction from measurement.

For the prior case (without measurement), the policy is applied to the prior probability model, and the expected utility it achieves over samples from the prior is evaluated,
\begin{equation} \label{eq:OP-VOI-prior-utility}
    \mathbb{E}_{\vartheta \sim \pi(\theta)}\left\lbrace u \left( \mathcal{P}_u \left( \pi(\theta) \right), \vartheta \right) \right\rbrace
\end{equation}

For the pre-posterior case (with uncertainty reduction from a measurement $e$), a set of hypothesised measurement values $z$ are sampled from the joint probability model, $\pi(\theta,z)$. For each hypothetical measurement, the posterior probability model $\pi(\theta|z)$ is constructed, the policy is applied to this posterior, and the expected utility it achieves over the posterior is computed. The pre-posterior expected utility is the expectation of these values over the hypothesised measurements,
\begin{equation} \label{eq:OP-VOI-pre-posterior-utility}
    \mathbb{E}_{z \sim \pi(\theta,z)} \left\lbrace \mathbb{E}_{\vartheta \sim \pi(\theta|z)}\left\lbrace u \left( \mathcal{P}_u \left( \pi(\theta|z) \right), \vartheta \right) \right\rbrace \right\rbrace
\end{equation}

Therefore the VoI for an imperfect measurement, $e$, used to support a policy, $\mathcal{P}_u$, is given by,
\begin{equation} \label{eq:OP-VOI}
    \text{EVII}_{\mathcal{P}}(e) = \mathbb{E}_{z \sim \pi(\theta,z)} \left\lbrace \mathbb{E}_{\vartheta \sim \pi(\theta|z)}\left\lbrace u \left( \mathcal{P}_u \left( \pi(\theta|z) \right), \vartheta \right) \right\rbrace \right\rbrace - \mathbb{E}_{\vartheta \sim \pi(\theta)}\left\lbrace u \left( \mathcal{P}_u \left( \pi(\theta) \right), \vartheta \right) \right\rbrace
\end{equation}

This framework extension is termed `On-Policy VoI', as the benefit of uncertainty reduction is now dependent on the policy. This broadens a main criticism of the VoI framework, namely that VoI results do not generalise and are only valid for the particular system and probability models used for calculations, as now results are additionally only valid for the particular policy. However, this also opens the opportunity for investigating which policies are most benefited by uncertainty reduction and best exploit additional statistical information.

The key advantage of `On-Policy VoI' is that it enables the study of the benefit of uncertainty reduction for complex, practical engineering decision problems where decision policies must be used, for instance problems with high-dimensional action spaces (i.e. where a large number of decisions must be made simultaneously), or those with tight computational limitations.
% ... can also be used to investigate deterministic policies, e.g. the Expected Value Problem \cite{leibowicz2018CostPolicyUncertainty} approach, which are commonly used in engineering applications ...

%% file: sections/experiment.tex
\section{Case study district energy system} \label{sec:experiment}

% Explain system setup and uncertainty reduction problem studied
% Provide intro chitchat explaining the problem setup & task - uncertain demand profiles & monitoring
% Mention grid constraints

The benefits of uncertainty reduction to support energy system design are investigated for a problem involving the sizing of solar-battery systems to decarbonize electricity usage in a grid-constrained district energy system with electrified heating. At the time of design, the patterns and scale of electrical load in the buildings of the district are unknown/uncertain. This load uncertainty must be managed during design to allow the district to operate within the grids constraints. Specifically, the peaks in electrical load must be manageable, though their magnitude and timing (particular the co-incidence of peaks) or not known during design.

Prior to committing to a system design (PV, battery, and grid connection sizes), the designer may choose to gather hourly electricity usage for the buildings from smart meters, reducing uncertainty in both the patterns and scale of their electrical load. This enables a more tailored design with a lower total cost to be identified. However, gathering hourly metering data for each building in the district and applying it to size the solar-battery systems is costly. This case study asks the question, 
\begin{itemize}
    \item[] ``Does building monitoring to reduce uncertainty in electrical load provide a net economic benefit when designing solar-battery systems?'',\\
    i.e. ``Does the reduction in cost from improved design outweigh the costs of collecting hourly monitoring data needed to achieve uncertainty reduction?''.
\end{itemize}

%\newpage
\subsection{Case study data} \label{sec:data}

% Historic data provides counterfactual case study (project results into future)
% Need to describe both Estates dataset and how it is processed - i.e. how heat electrification is modelled

Historic building energy usage data from the Cambridge University Estate is used to provide a case study for the design of a district of new academic buildings\footnote{At the time of writing, such a development project is under consideration, and one significant hurdle for its implementation is the capacity restrictions on the local power distribution network, particularly as heat pumps are a leading option for minimising heating related carbon emissions.}. The dataset, \cite{langtry2024CambridgeUniversityEstates}, contains hourly electricity and gas usage data from 121 buildings of various usage types from the University estate, such as lecture blocks, offices, laboratories, and museums, covering the period 2000 to 2023, as well as weather observations and grid electricity price and carbon intensity data. The electricity usage measurements record the total electrical load at a given metering point in the building, which includes lighting, plug loads, and plant equipment  electricity consumption. It is assumed that none of the buildings have heat pumps or AC units, so the electricity usage does not include any contributions from space heating or cooling. The gas usage measurements provide the total gas consumption at the metering points in each building, which is used for both space heating and DHW provision. Normalised power generation data from PV panels is taken from the \href{https://www.renewables.ninja/}{renewables.ninja} reanalysis model \cite{pfenninger2016LongtermPatternsEuropean,staffell2016UsingBiascorrectedReanalysis}, dynamic electricity pricing tariff data from Energy Stats UK \cite{energystatsuk2023HistoricalPricingData}, grid electricity carbon intensity from the National Grid ESO Data Portal \cite{nationalgrideso2020HistoricGenerationMix}. Further detail on the data sourcing and processing is available in reference \cite{langtry2024CambridgeUniversityEstates}.

\begin{figure}[b!]
    \centering
    \subfloat[Summer.]{
        \includegraphics[width=\linewidth]{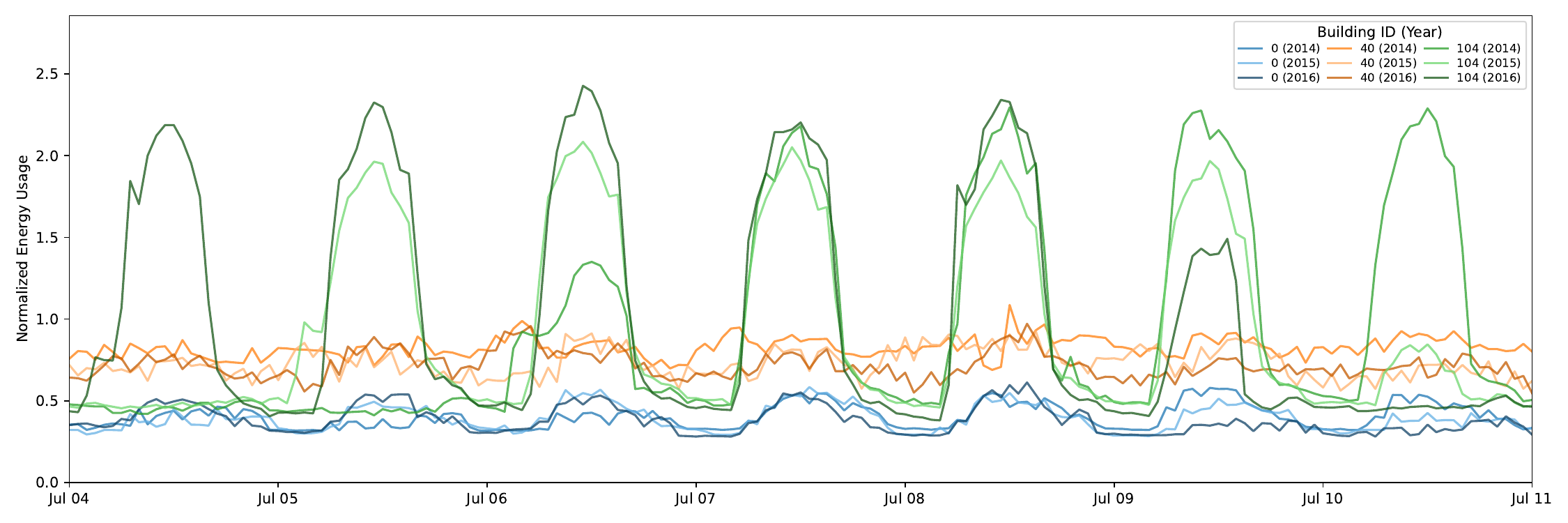} \label{fig:summer-loads}
    }

    \subfloat[Winter.]{
        \includegraphics[width=\linewidth]{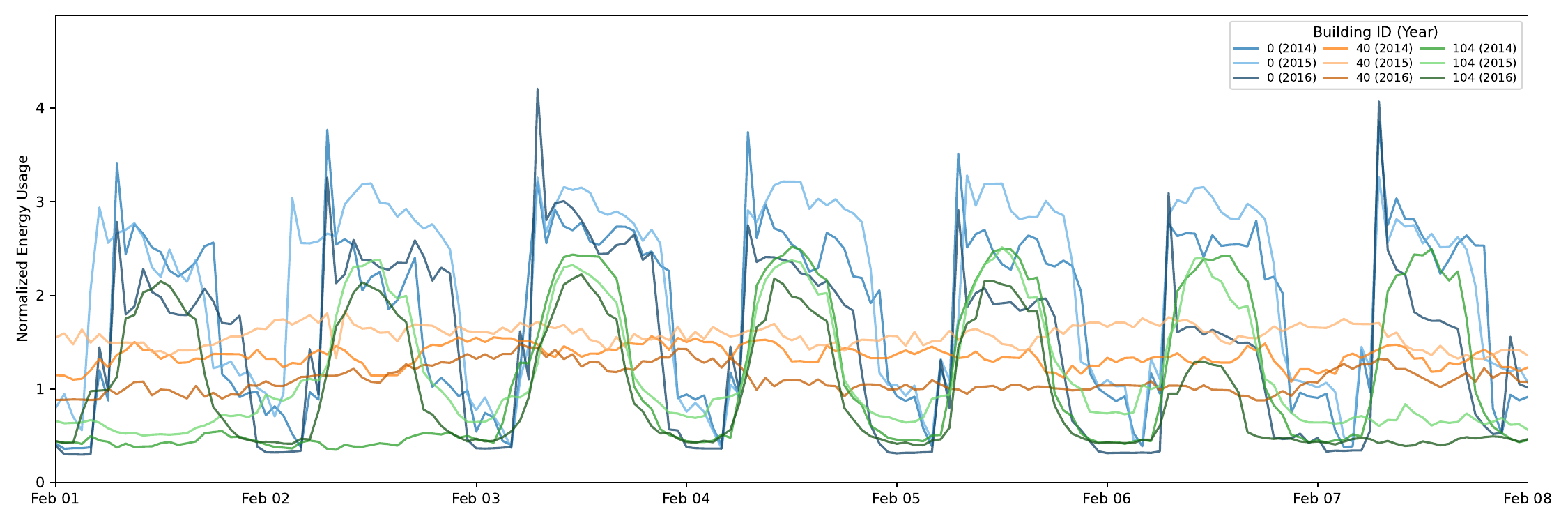} \label{fig:winter-loads}
    }
    \vspace*{-0.2cm}
    \caption{Examples of studied electricity load profiles, representing buildings with electrified heating.} \label{fig:example-load-data}
    \vspace{0.1cm}
    \small{\textit{The complete set of electrified building load profiles considered in the case study can be view interactively} \href{https://mal84emma.github.io/Building-Design-VoI/load_dataset_plot.html}{\textbf{here}}.}
\end{figure}

From this dataset, 9 buildings with available electricity and gas usage data for the period 2012 to 2017 are selected (the longest period in the dataset with a significant number of buildings with high quality data available). For each building, data for each year is separated out. To model energy usage in buildings with electrified heating, the electricity and gas usage measurements are combined to produce overall electrical load profiles such that equal amounts of electricity are used by direct electricity and heating (meaning that assuming a SCOP of 3, typical for ASHPs in the UK \cite{terry2023HowHeatDemand}, the direct electricity to final heating energy usage matches the current 25:75 ratio \cite{eea2023DecarbonisingHeatingCooling}). Fig. \ref{fig:example-load-data} plots examples of these building electrical load profiles (normalized by their mean loads) for periods in the summer and winter. It demonstrates the variation in electricity usage behaviour both between buildings, and between years for a given building.

\newpage
Data from 2023 is used for grid electricity price and grid carbon intensity, as these provide the closest available representation of the conditions the district will face during operation.
To represent the uncertainty in solar generation, years of normalised solar generation data are sampled randomly for each scenario from a dataset of the years 2010 to 2019.
For the case study, these data are treated as exogenous, and are used to defined the probabilistic operational conditions of the energy system.

\subsection{Probabilistic models of uncertain building electrical load \& measurement via monitoring} \label{sec:prob}

% Note that we are using data to model complex uncertainties in funcational variables (load profiles), which we parameterize
% Specifically looking at reduction in uncertainty in hourly load profile (important because of peaks), corresponding to data collected from building monitoring system
% Explain derivation of probabalistic models, may need to include figures (in appendix?)
% Highlight posterior representing *practical* measurement - derived from real data

A probabilistic model of the uncertain electricity load of a building is constructed from the data. The functional data variables representing building load are parameterized to provide a simple statistical representation that allows the importance of different characteristics of electrical load to be studied. Four parameters are used to specify the load profiles, two representing the shape of the profile (\texttt{type} and \texttt{year}), and two representing the scale of electricity usage (\texttt{mean} and \texttt{peak}).

The probabilistic models of profile shape are taken directly from data. The parameter \texttt{type} is used to represent the broad patterns of energy usage exhibited by a building, which derive largely from its usage type, e.g. laboratory, teaching, office, or a mix thereof. It is taken as a categorical variable denoting which building in the case study dataset the load profile data is taken from. All buildings are assumed equally likely,
\begin{equation} \label{eqn:id-model}
    \texttt{type} \sim \mathcal{U}\left(\lbrace \text{available buildings} \rbrace\right)
\end{equation}
The parameter \texttt{year} is used to represent the inherent randomness in building energy usage due to factors that cannot be known in advance, such as occupant behaviour and weather. It is taken as a categorical variable denoting the year from which the load profile data is taken, which are all assumed equally likely,
\begin{equation} \label{eqn:year-model}
    \texttt{year} \sim \mathcal{U}\left(\lbrace 2012,\ldots,2017 \rbrace\right)
\end{equation}
Load monitoring is assumed to provide knowledge of the broad energy usage patterns of a building, i.e. perfect information for the parameter \texttt{type} and no information for \texttt{year}.

Probabilistic models for the scale of electricity usage are determined from the statistics of the available load data. The prior model of mean building load is taken to be,
\begin{equation} \label{eqn:mean-model}
    \texttt{mean} \sim \mathcal{N}(100,25)
\end{equation}
representing a $\pm50\%$ uncertainty in a design load of 100kW.

The prior model of peak building load is taken to be,
\begin{equation} \label{eqn:peak-model}
    \texttt{peak} \sim \mathcal{U}(200,400)
\end{equation}
which is fit using the annual peak loads from the dataset.

In a practical building, load monitoring provides an uncertain/imperfect estimate of the mean and peak building load, as there is year-on-year variation in these values that cannot be predicted in advance. The likelihood models for the precision of mean and peak load measurements (estimates) from building monitoring are taken to be Gaussian, with measurement errors determined using the year-on-year variations observed in the load dataset, i.e. the likelihood of obtaining a measurement value is,
\begin{equation}
    z|\theta \sim \mathcal{N}(\theta,\varepsilon\theta)
\end{equation}
with $\varepsilon = 0.1$ for \texttt{mean}, and $\varepsilon = 0.075$ for \texttt{peak}.\\

The prior probabilistic models of the parameters describing the building electricity load profiles, and the likelihood models for measurements made via building monitoring are summarized in Table \ref{tab:prob-models}. Where necessary, the corresponding posterior distributions are modelled using \texttt{Stan} \cite{carpenter2017StanProbabilisticProgramming}.

The probabilistic models of electrical load are taken to be the same for all buildings in the district, with the parameters \texttt{type}, \texttt{mean}, and \texttt{peak} independent for each building. The \texttt{year} parameter is the common to all buildings (sampled once for the district in each scenario). This is done to ensure that correlations in the timing of energy usage between buildings are properly captured, as building energy usage is driven by common externalities across the district, for example outdoor temperature which affects heating load. These correlations in building loads are important for correctly representing the overall energy usage behaviour of the district. For example, the relative timing of peaks in building load determine the district peak load, and so the required grid connection capacity.

Samples are drawn from the building load distributions (either prior or posterior) by sampling the four parameters (\texttt{type}, \texttt{year}, \texttt{mean}, \texttt{peak}) using their probabilistic models, retrieving the constructed electrified heating load data for the sampled building (\texttt{type}) and year (see Fig. \ref{fig:example-load-data}), and scaling the profile to the sampled mean and peak load values.

\begin{table}[h]
    \centering
    \renewcommand{\arraystretch}{1.25}
    \begin{tabular}{c|cccc} \toprule \toprule
        Parameter & Prior model & Prior params & Measurement model & Msr. params \\
        \midrule \midrule
        \texttt{type} & \multirow{2}{*}{Discrete uniform} & $\{ \text{building ids} \}$ & Exact (perfect info) & -- \\
        \texttt{year}* & & $\{ \text{data years} \}$ & None (no info) & -- \\
        \midrule
        \texttt{mean} & Gaussian & \makecell{$\mu=\SI{100}{kW}$\\$\sigma=\SI{25}{kW}$} & \multirow{2}{*}{\makecell{Gaussian imprecision\\(imperfect info)\\[1ex]$z|\theta \sim \mathcal{N}(\theta,\varepsilon\theta)$}} & $\varepsilon = 0.1$ \\[2ex]
        \texttt{peak} & Continuous uniform & \makecell{$\min=\SI{200}{kW}$\\$\max=\SI{400}{kW}$} & & $\varepsilon = 0.075$ \\
        \bottomrule \bottomrule
    \end{tabular}
    \caption{Probabilistic models of parameters describing uncertain building load, and their measurement via monitoring.}
    %\vspace*{-0.1cm}
    {\footnotesize *\textit{Parameter common to all buildings in the district}}
    \label{tab:prob-models}
\end{table}

\newpage

The prior probabilistic model of building electrical load is visualised in Fig. \ref{fig:load-dist}. The mean load profile is indicated in black, with confidence intervals of $\pm 1 \text{ s.d.}$, $\pm 2 \text{ s.d.}$, and the range ($\min$ to $\max$), indicated in successively lighter shades of gray. A set of load profiles sampled from the distribution are traced in blue.

Fig. \ref{fig:post-load-dist} illustrates the posterior distribution of building load for an example load profile measurement obtained by building monitoring (indicated in red). Load profiles sampled from the posterior distribution are traced in green. The confidence intervals of the prior distribution, from Fig. \ref{fig:load-dist}, are underlaid for comparison. The tighter clustering of the profiles sampled from the posterior compared to that for the prior samples in Fig. \ref{fig:load-dist} demonstrates the reduction in building load uncertainty achieved by monitoring.\\

\newcommand{\tsfw}{1}
\begin{figure}[!h]
    \centering
    \subfloat[Summer.]{
        \includegraphics[width=\tsfw\linewidth]{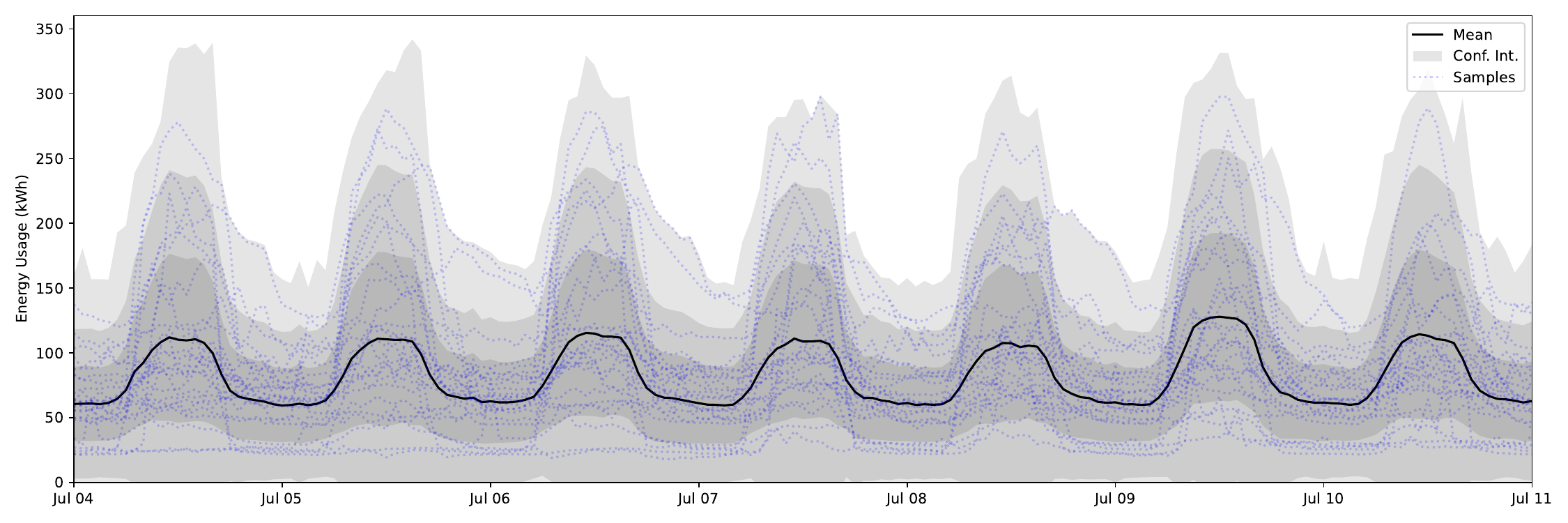} \label{fig:summer-dist}
    }

    \subfloat[Winter.]{
        \includegraphics[width=\tsfw\linewidth]{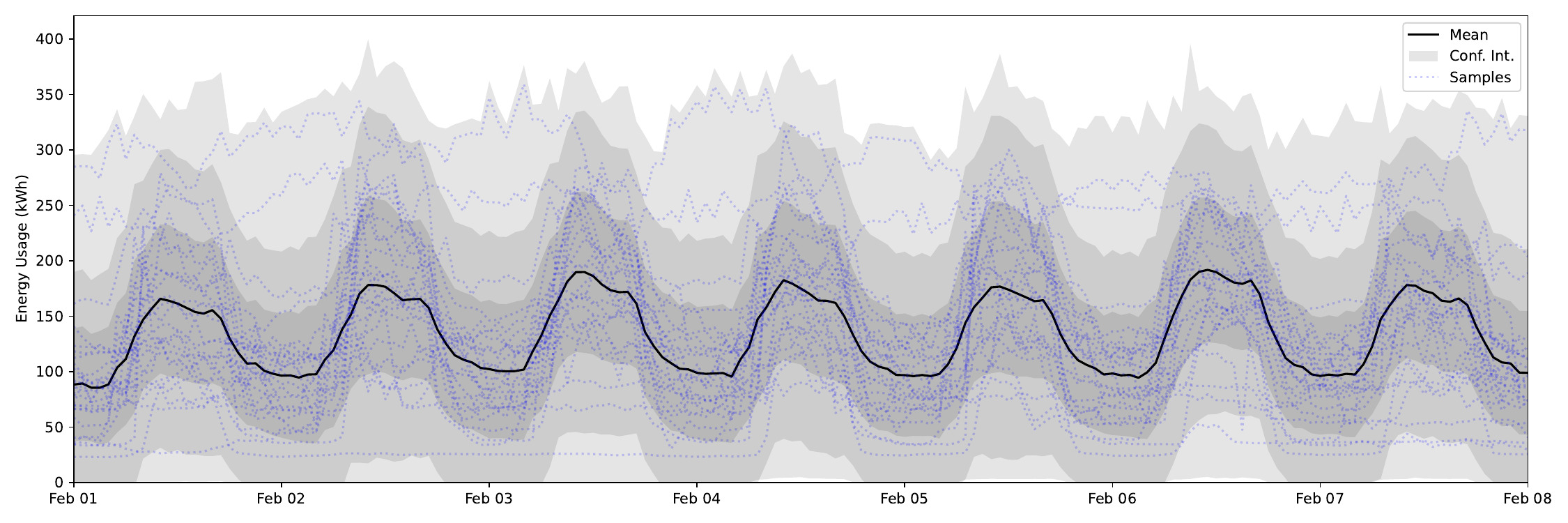}
        \label{fig:winter-dist}
    }
    \vspace*{-0.2cm}
    \caption{Visualisation of prior probabilistic model of uncertain building load, with profiles sampled from distribution.} \label{fig:load-dist}
\end{figure}

\begin{figure}[ht]
    \centering
    \subfloat[Summer.]{
        \includegraphics[width=\tsfw\linewidth]{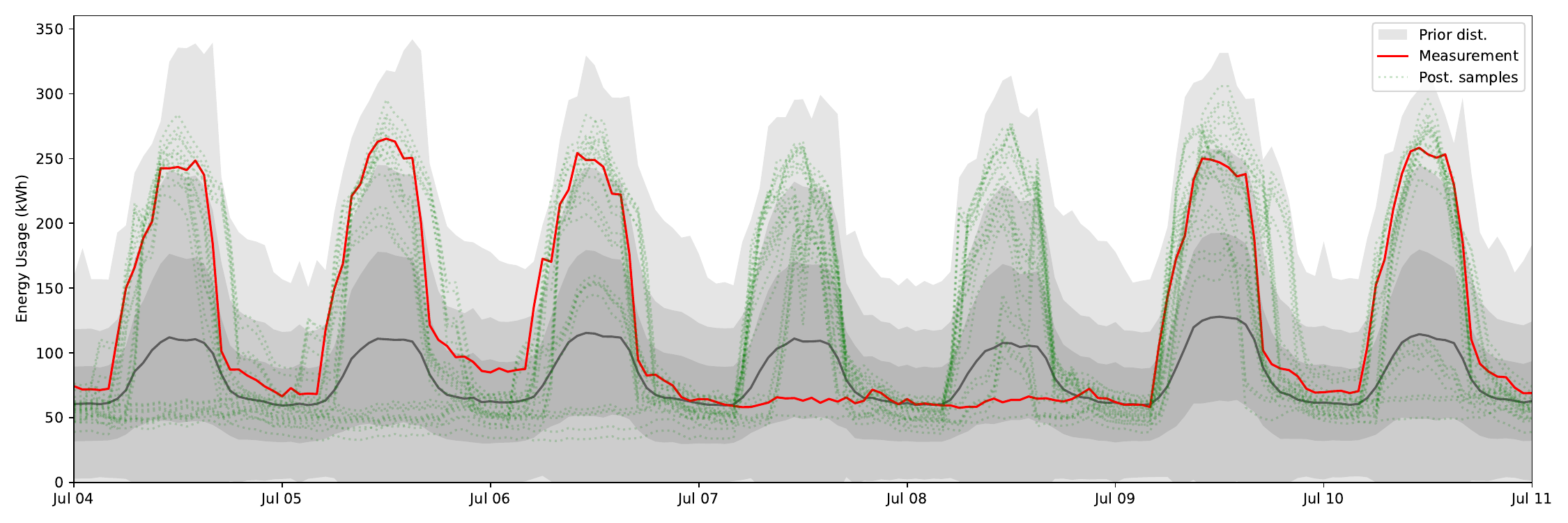} \label{fig:summer-post-dist}
    }

    \subfloat[Winter.]{
    \includegraphics[width=\tsfw\linewidth]{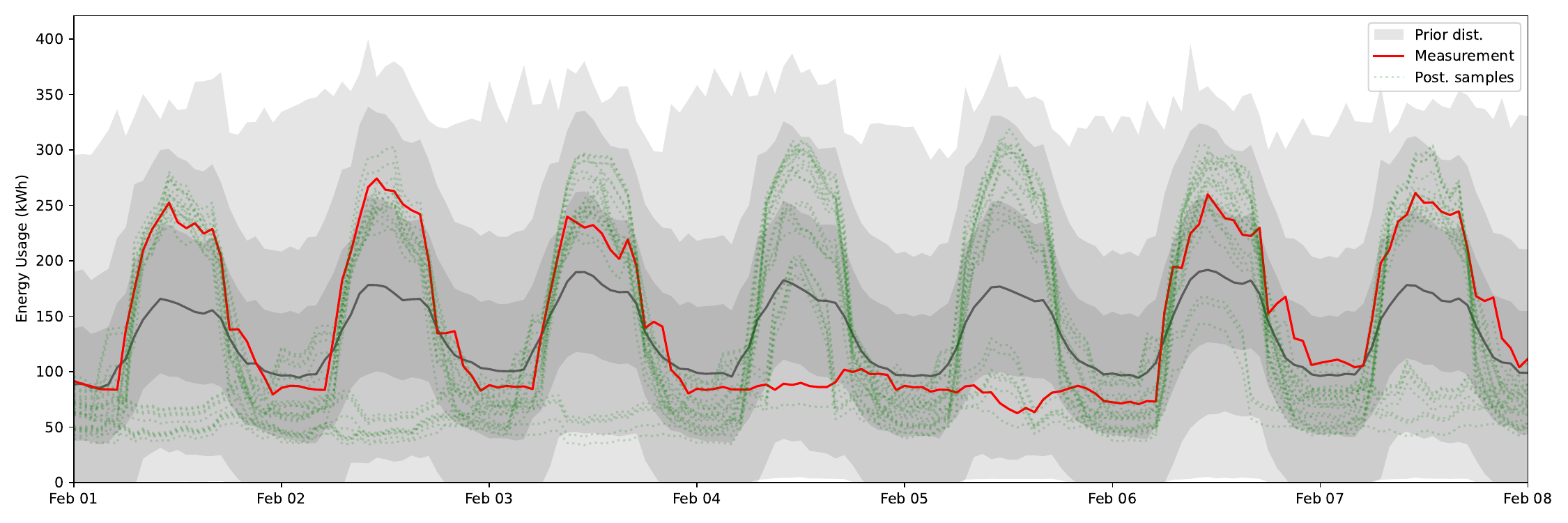}
    \label{fig:winter-post-dist}
    }
    \caption{Visualisation of posterior probabilistic model of building load for example measurement.} \label{fig:post-load-dist}
\end{figure}

\newpage
\subsection{Stochastic Programming for system sizing} \label{sec:SP}

% mention why Stochastic Programming chosen for system design (popular choice and well suited for systems with lots of decision variables - e.g. district energy systems)
% remember to discuss scenario reduction - need to use simple metrics (instead of op. results) as scenario space is combinatorial (cite Bryn's thesis)
% explain form of costs (grid connection costs from DUoS charges)
% comment that SP does not provide accurate estimate of true system cost due to: safety factor, over-optimism induced by excessive information availability (leading to need for safety factor)

For the case study, Stochastic Programming is used as a policy to optimize the sizing of solar PV, battery storage units, and grid connection capacity, to determine district energy system designs that (attempt to) minimize the expected total system cost, accounting for uncertainty in electrical load and solar generation. Stochastic Programming (specifically linear scenario programming) is chosen as the modelling and optimization methodology due to its computational efficiency for problems with large numbers of decision variables, and its resulting prevalent use in the current literature for designing district- and national-scale energy systems \cite{decarolis2017FormalizingBestPractice,pickering2019DistrictEnergySystem,yue2018ReviewApproachesUncertainty,2024OpenEnergyModelling,yang2015MILPMixedInteger}.

\begin{figure}[h]
    \centering
    \includegraphics[height=6.5cm]{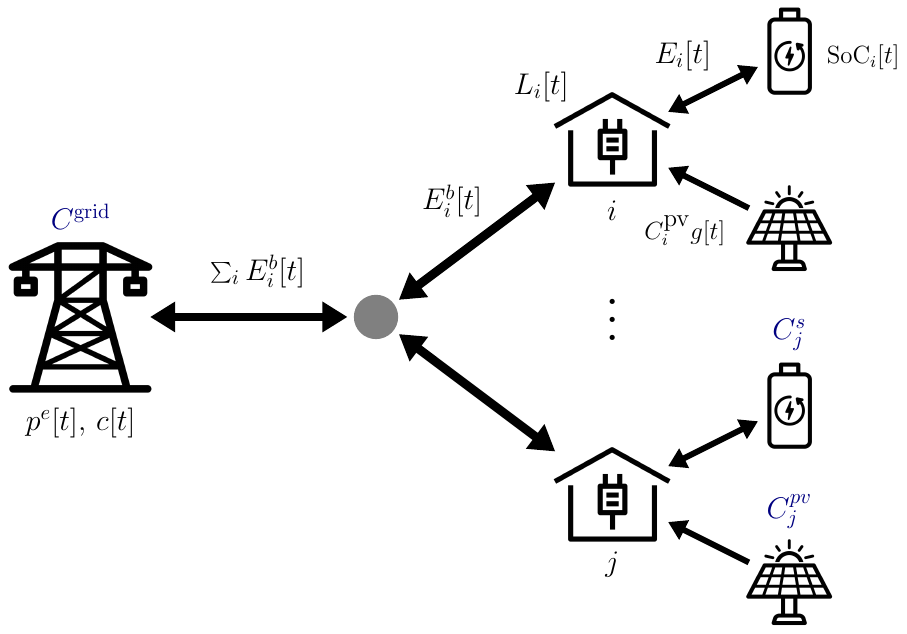}
    \caption{Energy flow schematic for case study district energy system.}
    \vspace*{-1ex}
    {\scriptsize Operational variables are indicated in black, capacity variables are indicated in blue. (Icon credits: \href{https://thenounproject.com/symbolon/}{Symbolon})}
    \label{fig:energy-system}
\end{figure}

A linearized model of the electrical behaviour of the district energy system is used. A schematic of the energy flows within the district energy system model is provided in Fig. \ref{fig:energy-system}. This allows the design optimization task to be formulated as a Linear Program, which is presented in Eq. \ref{eq:SP}, with Table \ref{tab:LP-params} providing descriptions of the parameters. The model formulation uses the same form of constraints to represent energy system behaviour as state-of-the-art models such as \cite{pfenninger2018CalliopeMultiscaleEnergy} and \cite{brown2018PyPSAPythonPower}.

%\newpage

\newcommand{\eqnskip}{1ex}
\newcommand{\smalleqnskip}{-1ex}
\begin{subequations} \label{eq:SP}
    \begin{align}
        \addtocounter{equation}{-1}
        \begin{split}
        \mathlarger{\min} & \qquad \mathlarger{\gamma} \:\: \mathlarger{\mathlarger{\sum}}_m \: \rho_m \vast( \mathlarger{\sum}_t \Bigg( \underbrace{p^e[t] \sum_i \left( \max \left[ 0,\, E_{i,m}^b[t] \right] \right)}_{\text{electricity cost}} + \, \underbrace{p^c \, c[t] \, \max \Bigg[ 0,\, \sum_i E_{i,m}^b[t] \Bigg]}_{\text{carbon cost}} \Bigg) \\
        & \hspace{3cm} + \:\: \underbrace{p^{\textrm{grid}}_{\textrm{excess}} \max \Bigg[0, \max_t \bigg[ \Big| \sum_i E_{i,m}^b[t] \Big| \bigg]\Big/\Delta t - C^{\textrm{grid}}/\text{FoS} \Bigg]}_{\text{grid excess cost}} \vast)\\
        & \hspace{1.5cm} + \underbrace{\mathlarger{\mathlarger{\sum}}_i \bigg( p^{\textrm{s}} C^{\textrm{s}}_i + p^{\textrm{pv}} C^{\textrm{pv}}_i \bigg)}_{\text{capital cost}} + \underbrace{\vphantom{\mathlarger{\sum}_i} \mathlarger{\gamma} \: p^{\textrm{grid}} C^{\textrm{grid}} \:}_{\text{grid connection cost}}
        \end{split} \label{eq:lp} \\[\eqnskip]%%
        \text{over} & \qquad C^{\textrm{s}}_i,\, C^{\textrm{pv}}_i,\, C^{\textrm{grid}},\, E_{i,m}[t],\, \textrm{SoC}_{i,m}[t{+}1] \quad \forall \: i,\, m,\, t \tag*{} \\[\eqnskip]
        \text{subject to} & \qquad \textrm{SoC}_{i,m}[t{+}1] = \textrm{SoC}_{i,m}[t] + \sqrt{\eta_i} \left[E_{i,m}[t]\right]^{+} - 1/\sqrt{\eta_i} \left[\, E_{i,m}[t] \right]^{-} \label{eq:dynamics-constraint} \\[\smalleqnskip]
        & \qquad -P^{\textrm{max}}_i \Delta t \leq E_{i,m}[t] \leq P^{\textrm{max}}_i \Delta t \label{eq:power-constraint} \\[\smalleqnskip]
        & \qquad 0 \leq \textrm{SoC}_{i,m}[t{+}1] \leq C^{\textrm{s}}_i \label{eq:energy-constraint} \\[\smalleqnskip]
        & \qquad C^{\textrm{s}}_i,\, C^{\textrm{pv}}_i,\, C^{\textrm{grid}} \geq 0 \\[\smalleqnskip]
        & \qquad P^{\textrm{max}}_i = \delta C^{\textrm{s}}_i \label{eq:power-capacity} \\[\smalleqnskip]
        & \qquad \textrm{SoC}_{i,m}[0] = \textrm{SoC}^0 C^{\textrm{s}}_i \label{eq:initial-conditions} \\[\smalleqnskip]
        & \qquad E^b_{i,m}[t] = L_{i,m}[t] - C^{\textrm{pv}}_i g_m[t] + E_{i,m}[t] \label{eq:aggregation-constraint} \\[\eqnskip]%%
        \text{for all} & \qquad i \in [0,B{-}1], \: t \in [0,T{-}1], \: m \in [0,M{-}1] \tag*{}
    \end{align}
\end{subequations}

\begin{table}[h]
    \centering
    \renewcommand{\arraystretch}{1}
    \begin{tabularx}{\linewidth}{ccX} \toprule \toprule
        Parameter & Units & \multicolumn{1}{>{\centering\arraybackslash}c}{Description} \\
        \midrule \midrule
        \multicolumn{3}{>{\centering\arraybackslash}l}{\small \quad Decision variables} \\
        $C^{\textrm{s}}_i$ & kWh & Energy capacity of battery unit in building $i$ \\
        $C^{\textrm{pv}}_i$ & kWp & Peak power capacity of solar PV unit in building $i$ \\
        $C^{\text{grid}}$ & kW & Contracted power capacity of grid connection \\
        $E_{i,m}[t]$ & kWh & Energy \textit{intake} to battery unit in building $i$ at time $t$ in scenario $m$ \\
        $\textrm{SoC}_{i,m}[t]$ & kWh & State-of-charge of battery unit in building $i$ at time $t$ in scenario $m$ \\
        \midrule
        \multicolumn{3}{>{\centering\arraybackslash}l}{\small \quad Derived variables} \\
        $E^b_{i,m}[t]$ & kWh & Net energy flow \textit{into} building $i$ at time $t$ in scenario $m$ \\
        $P^{\textrm{max}}_i$ & kW & Power capacity of battery unit in building $i$ \\
        \midrule
        \multicolumn{3}{>{\centering\arraybackslash}l}{\small \quad Parameters (data)} \\
        $\gamma$ & yrs & System lifetime \\ %(multiple of operational period, $T$) \\
        $\Delta t$ & hrs & Time step of simulation data \\
        $\rho_m$ & -- & Probability of scenario $m$ \\
        $\delta$ & kW/kWh & Discharge ratio of battery units (power capacity/energy capacity) \\
        $\textrm{SoC}^0$ & -- &  Initial state-of-charge of battery units (fraction of capacity) \\
        $\eta_i$ & -- & Round-trip efficiency of battery unit in building $i$ \\
        $L_{i,m}[t]$ & kWh & Electrical demand of building $i$ at time $t$ in scenario $m$ \\
        $g_m[t]$ & kW/kWp & Normalised generation power from solar PV at time $t$ in scenario $m$ \\
        $p^e[t]$ & £/kWh & Grid electricity price at time $t$ \\
        $p^c$ & £/kgCO$_2$ & Nominal carbon price \\
        $c[t]$ & kgCO$_2$/kWh & Carbon intensity of grid electricity at time $t$ \\
        $p^{\textrm{s}}$ & £/kWh & Unit price of battery storage \\
        $p^{\textrm{pv}}$ & £/kWp & Unit price of solar PV \\
        $p^{\text{grid}}$ & £/kW & Price of contracted grid connection capacity (over operational period) \\
        $p^{\text{grid}}_{\text{excess}}$ & £/kW & Excess charge for exceeding contracted grid capacity (over op. period) \\
        $\text{FoS}$ & -- & Factor of safety for grid connection capacity \\
        \bottomrule \bottomrule
    \end{tabularx}
    \caption{Description of Stochastic Program variables \& parameters.} \label{tab:LP-params}
\end{table}
\hfill

The design optimization seeks to minimize as its objective the average total lifetime system cost over a set of $M$ scenarios for the district of $B$ buildings, operated for period of length $T$. This total system cost is comprised of: the capital cost of solar PV and battery units in the buildings, the cost of electricity purchased from the grid (metered per building), the cost of carbon emissions associated with used grid electricity (for the overall district), and the cost of grid usage.
This optimization is subject to energy conservation\footnote{$[\,\cdot\,]^+$ and $[\,\cdot\,]^-$ represent the positive and negative parts of the argument respectively.} (Eq. \ref{eq:dynamics-constraint}) and capacity (Eq. \ref{eq:power-constraint} \& Eq. \ref{eq:energy-constraint}) constraints for operation within each scenario $m$.

The cost of grid usage is modelled following the form of Distributed Use of Service (DUoS) connection capacity charges, which are the costs charged to large electricity consumers connected to the distribution network in the UK \cite{acha2022ModellingUKElectricity,frontiereconomics2022NetworkTariffsEnergy}. These consumers must request a `contracted' connection capacity from the distribution operator, for which they are charged a cost per unit of connection capacity (£/kVA). If during operation a consumer draws power from the grid exceeding this contracted capacity, they are charged an additional cost for each kVA that their peak power draw exceeds the contracted capacity. Exaggerated grid charges (compared to current values) are used to represent the pressure on network connection capacity caused by future heat electrification.

As Linear Programs simultaneously optimize over all operational variables, they assume that the district energy system is operated with perfect foresight. This leads to underestimation of the system operational costs, as the battery units within the modelled energy system are scheduled more effectively than could be achieved in practice, reacting to and avoiding future costs which could not be foreseen. This is particularly important for the grid excess costs which depend on peak grid load values that are greatly influenced by the perfect foresight assumption. To account for this underestimation of grid usage in the design model, a factor of safety ($\text{FoS}$) is introduced into the grid excess cost. This $\text{FoS}$ was calibrated using preliminary experiments, and was found to greatly reduce the overall cost of designs, as evaluated using the operational simulations introduced in the following section. % can be easily shown if reviewer asks

The Stochastic Program requires samples (scenarios) to be drawn from the probability space of electrical load profiles for each building in the district. As a result, the space of possible load scenarios is extremely large. However, due to the cubic computational complexity of Linear Programming, only a small number of scenarios can be tractably included in the Scenario Optimization. Therefore, after load profile samples are drawn, scenario reduction \cite{heitsch2003ScenarioReductionAlgorithms,gioia2023ScenarioReducer} is performed to select a reduced set of 10 scenarios\footnote{The maximum number which could be used by the Stochastic Program given the available computational budget.} that are statistically representative of the initial sample. A large initial sample must be considered to adequately cover the probability space of district loads. Therefore state-of-the-art scenario reduction methods which optimize each scenario individually, and use the resulting distribution of costs as the reduction metric \cite{pickering2019PracticalOptimisationDistrict} are intractable. Instead, computationally cheap scenario metrics must be used. The mean, maximum, and standard deviation of the aggregate electrical load for the district are chosen as the metrics for scenario reduction, as simple features representing the shape and scale of the district electricity load.

The parameter values used for all experiments are detailed in \ref{app:case-study-params}.

\subsection{Evaluation of operational cost of system designs via simulation} \label{sec:sims}

% Explain why we want to use the simulation for evals (SP doesn't accurately estimate cost, due to; over-optimism, small sample size) - highlight design vs operation distinction from SP
% Receeding horizon MPC using same LP as design (reduced FoS), over horizon $\tau$ with perfect forecasts
% Generally SP objectives are not fully reliable (under-estimate costs due to excessive info availability), but in this case adding in the heuristics improves system performance, but means the SP objective is not the actual system cost, so additional eval needed anyway
% Highlight importance of design LP and simulations being mathematically identical (MPC model matches simulator), with the only difference being the quantity of information available for making control decision (omniscience vs finite horizon) - and the objective fn

The objective of the Stochastic Program (SP) provides only an approximation of the expected (mean) total cost of the district energy system. The perfect foresight assumption of the Linear Program model causes underestimation of operating costs, a modelling error which is only partly corrected by the factor of safety. The small number of scenarios which can be considered in the Scenario Optimization due to computational limitations causes statistical error in the estimation of the mean cost.
The combined model and statistical errors could lead to significant inaccuracy in the cost estimates\footnotemark, and so the VoI. To provide more accurate estimations of the true cost of system designs (the decision maker's target utility), the operation of the district energy system is simulated.
\footnotetext{The accuracy of Stochastic Program objective's estimate of total system cost is investigated in \ref{app:cost-accuracy}.}

The CityLearn \cite{vazquez-canteli2019CityLearnV1OpenAI,vazquez-canteli2020CityLearnStandardizingResearch} building energy control framework is used to simulate the behaviour of the district energy system. These simulations provide a more accurate estimate of the operational cost of a system design for a given simulated load scenario. A Model Predictive Control (MPC) scheme is used to schedule battery operation in the simulations. This MPC scheme uses a Linear Programming (LP) formulation with the same form as used for design (Eq. \ref{eq:SP}), however it considers only a single scenario (that being simulated) with perfect forecasts of operational conditions over a 48hr planning horizon. Additionally, a small $\text{FoS}$ is used to prevent numerical errors from the LP solver leading to erroneous exceedance of the grid capacity, calibrated from initial experiments. The simulator is configured with linearized building dynamics, such that the MPC scheme perfectly models the system simulation, and so provides perfect system control with 48hrs of foresight, mirroring the SP used for design, but relaxing the perfect foresight assumption.
As a result, the simulations still underestimate operational costs \cite{langtry2024ImpactDataForecasting}, but this model error is lower than for the SP. As the simulations are computationally cheap, a large number of load scenarios can be simulated, and so the expected cost can be estimated using a large number of Monte Carlo samples (scenarios sampled from the building load distribution), reducing statistical error.

\subsection{Decision problem formulation}

% Briefly but clearly contextualise setup in terms of On-Policy VoI framework
% probabilistic models of uncertain load (pi(theta)) and monitoring measurements (f(z|theta)), SP defines policy, simulations define utility fn - got through section by section and link back to eqn. \ref{eq:OP-VOI}

The case study can be contextualized within the On-Policy VoI framework proposed in Sec. \ref{sec:on-policy} as follows.

\begin{figure}[p]
    \centering
    \includegraphics[width=\linewidth]{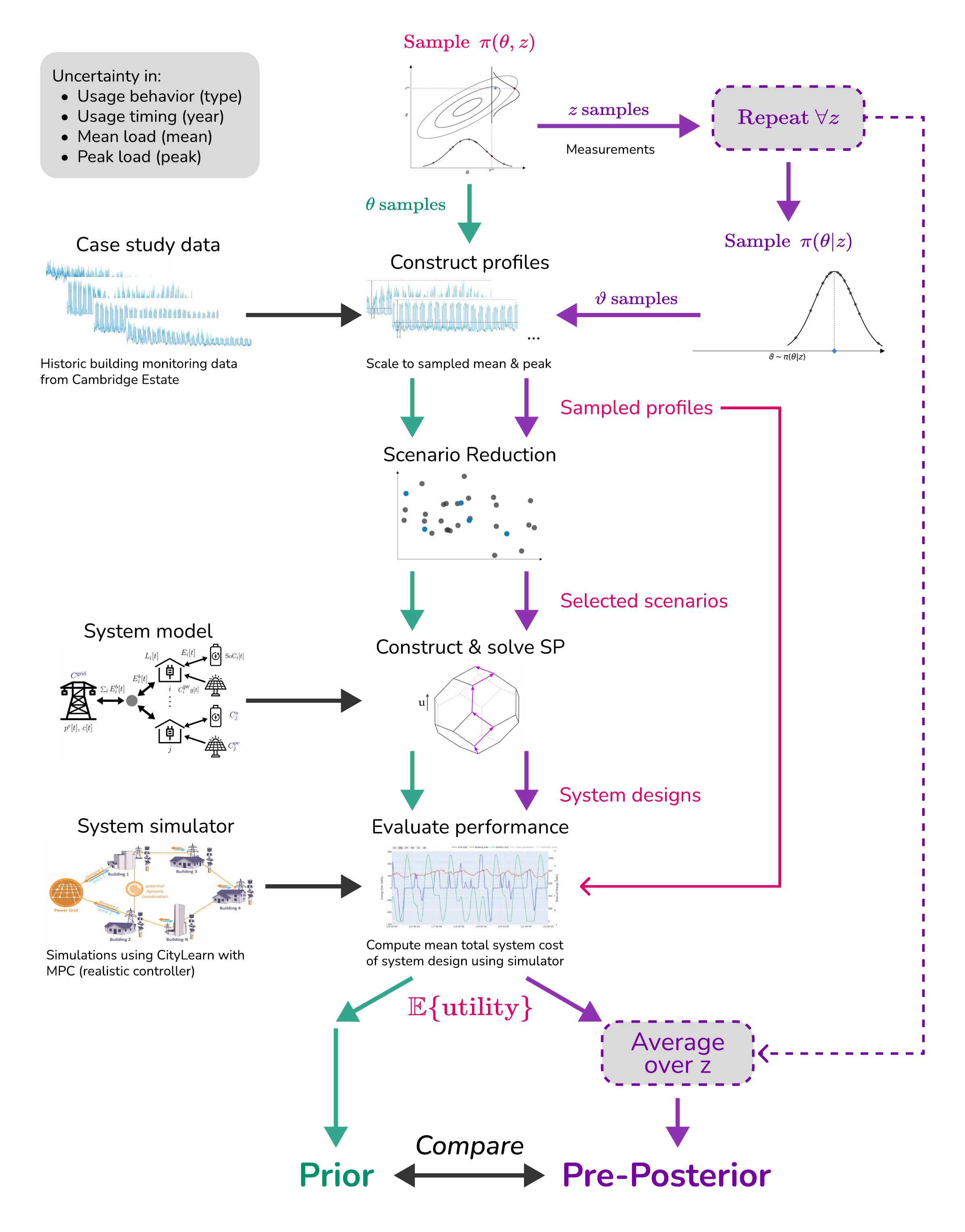}
    \caption{Methodology for computing VoI for case study district energy system design task.}
    \label{fig:method}
\end{figure}

The uncertain parameter in the district energy system, $\theta$, is defined as the electrical load profile (functional variable) in each building, and its probabilistic model, $\pi(\theta)$, is described in parametric form in Sec. \ref{sec:prob} (see Table \ref{tab:prob-models}). The likelihood function, $f(z|\theta)$, describing the uncertainty reduction (imperfect parameter estimates) obtained from building load monitoring, is also described.
A Stochastic Programming model, outlined in Sec. \ref{sec:SP}, is used as a policy, $\mathcal{P}_u$, determining a system design (action) to implement given samples from a distribution of building load profiles (to which scenario reduction is applied due to computational limitations). This policy minimizes an approximation of the expected total system cost (utility). The true system cost (utility, $u$) for a given design is evaluated using simulations of the district energy system with a more realistic control scheme, described in Sec. \ref{sec:sims}. Fig. \ref{fig:method} provides a diagrammatic overview of the methodology.

%% file: sections/results.tex
\section{Results \& Discussion} \label{sec:results}

\subsection{System design without load monitoring} \label{sec:prior-design}

%%! Think about narrative (VoI, load monitoring). What is relevant to discuss here?

%% !Briefly! explain and present initial (prior) design
% Discuss cost reduction provided by installing solar battery system - compare to costs without, with only solar, with only battery - *actually* is this part of the story? maybe just briefly mention the % cost reduction provided by the solar-battery system (and maybe specifically carbon emissions contribution also)
% No grid capacity exceedance (in 1000 samples) - design & controller about to manage system properly (important for context, we can have confidence in requested grid capacity) - grid impact can be limited to known/contracted value in all(?) cases

% Discuss impact of uncertainties on simulation outcomes for prior design - show that they are signficant, demonstrate that there is variation due to load profiles (uncertainty), and discuss contributions (mean most influential, shape signficant, peak uncorrelated)
% There is significant variability in cost outcomes due to load uncertainties which *could* be addressed/reduce high costs through measurement

A district containing 5 university buildings, without constraints on asset sizing, is considered as a base case. Initially the case where no building load monitoring is used is considered. Load profile scenarios were sampled from the prior distribution, which represents the behaviour of a generic university building, and the Stochastic Program model described in Sec. \ref{sec:SP} was used to size the solar-battery systems and grid connection capacity using these prior scenarios.

The optimized (prior) design installs 4,910 kWh of batteries and 2,790 kWp of solar panels across the 5 buildings, roughly evenly distributed, and contracts 1,230 kW of grid connection capacity (2.5 times the mean electrical load of the district of 500 kW). From operational simulations, the average lifetime cost for the system (the prior expected utility, Eqn. \ref{eq:OP-VOI-prior-utility}) was determined to be £22.432m over a 20 year operational lifetime, or £1.122m/year. Comparing to the case where no solar-battery systems are installed in the buildings, which results in a mean operational cost of £1.839m/year, this investment in local generation and storage reduces total operating costs for the energy system by 39\%. Additionally, the embodied carbon emissions of providing energy are reduced by 64\%, and the required grid connection capacity is reduced by 27\%. Further, the practical building controller did not exceed the contracted grid capacity in any of the 256 sampled scenarios simulated. These results validate the ability of solar-battery systems to simultaneously reduce the overall cost of energy provision, embodied carbon emissions, and the grid impact of the district. Further details of the optimized system design and its performance across the simulated scenarios are provided in \ref{app:prior-design}.

\begin{figure}
    \centering
    \subfloat[Total system costs.]{
        \includegraphics[width=0.475\linewidth]{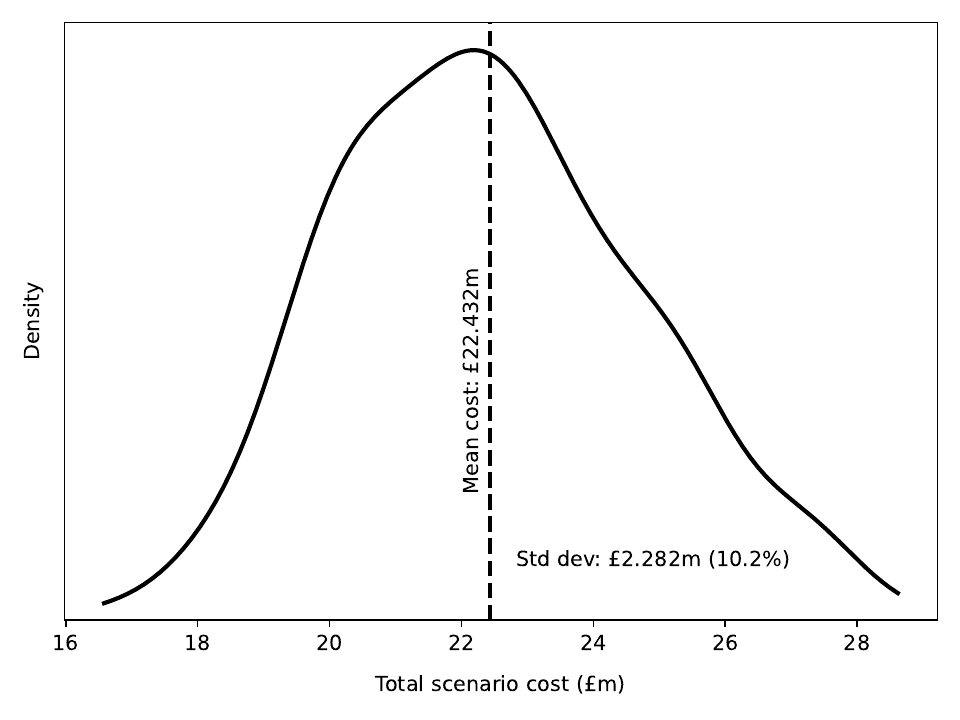} \label{fig:prior-costs-total}
    }%
    \hfill
    \subfloat[LCOEs.]{
        \includegraphics[width=0.475\linewidth]{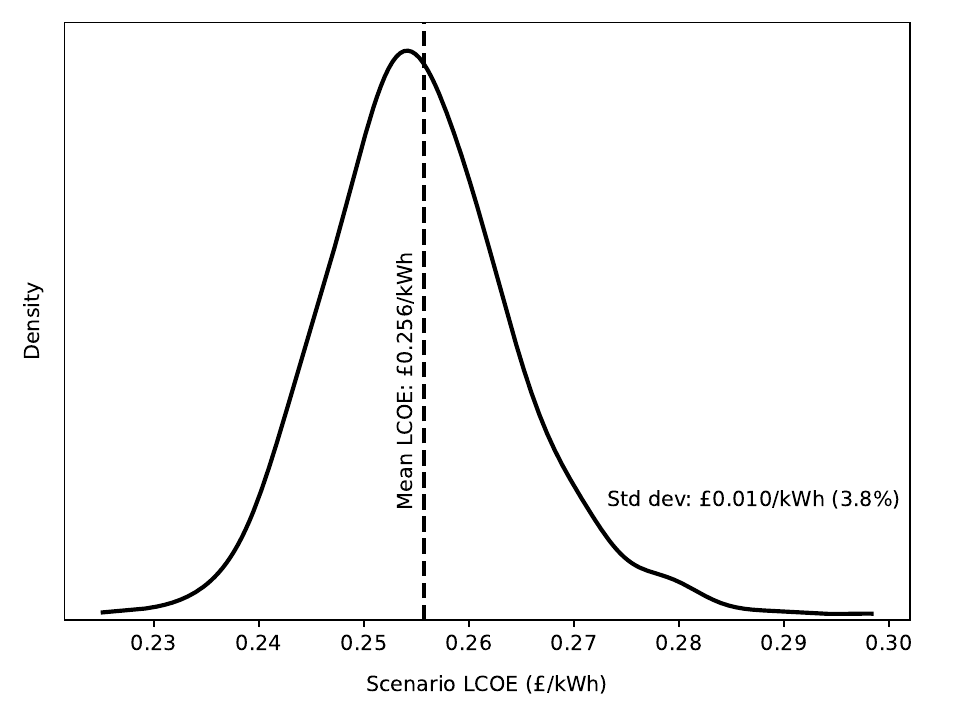} \label{fig:prior-costs-lcoe}
    }
    \vspace*{-0.2cm}
    \caption{Distribution of simulated total system costs and LCOEs for prior system design over sampled building load scenarios.} \label{fig:prior-costs}
\end{figure}

Fig. \ref{fig:prior-costs} plots the distributions of simulated system costs and corresponding Levelized Costs of Energy (LCOEs) when the prior design is operated for each sampled load scenario. It shows that the uncertainty in building load causes significant uncertainty in the total operating cost of the system. As the LCOE normalizes the operating cost by the total energy usage of the buildings, it removes the contribution of mean load to cost variability. The standard deviation of the LCOE distribution is 3.8\%, compared to 10.2\% for the total cost distribution, comparing Fig. \ref{fig:prior-costs-lcoe} to Fig. \ref{fig:prior-costs-total}. This demonstrates that mean load is the greatest driver of cost variability, but that the remaining uncertainties (peak load and load profile shape) also make a significant contribution.
There is found to be no correlation between the peak electrical load of the district and the total operating cost across all scenarios, indicating that peak load has little to no influence on the cost of operation. This can be justified as the sizing of the battery units to perform bulk energy arbitrage results in them having a power capacity of 1,960 kW, meaning they can handle the highest possible peak load of the district of 2000 kW (of which only 770 kW needs to be shaved to avoid exceeding the grid connection capacity). As a result, in almost all simulated scenarios no grid excess charges are incurred. Though the peak may occur when electricity is expensive, as they are infrequent there is only a small impact on the overall electricity cost. The operational behaviour of the district energy system and the factors influencing system design are discussed further in \ref{app:prior-design}.

This analysis demonstrates that uncertainty in mean electrical load of the district is the largest driver of operational cost uncertainty, responsible for over half of induced variability, but that the other uncertainties (in peak load, load profile shape, and solar generation) also has a significant influence on cost.
% Add interpretation comment? E.g. "This is because ..."
Therefore there might exist scope to reduce the total system cost by reducing uncertainty in these factors, and tailoring the system design to the energy usage behaviour of the true target building\footnotemark.
% So VoI study is meaningful (bit more naunce, but good enough)
% Explicitly state existing methods stop here and discuss weakness of conclusion provided
Most existing analyses, such as those in \cite{mavromatidis2018ReviewUncertaintyCharacterisation,prataviera2022EvaluationImpactInput}, or other purely statistical approaches \cite{pang2020RoleSensitivityAnalysis}, stop at this stage and put forward the conclusion that reducing uncertainty in the models inputs (here, building load) would significantly reduce uncertainty in the model output (the operating cost of the system). However, this outcome does not provide any indication about whether this uncertainty reduction would affect the decision made, and so cannot quantify its importance for supporting decision making.
\footnotetext{This tailoring of the system (improved design) could reduce not only the highest cost scenarios for the prior case (the right tail of Fig. \ref{fig:prior-costs-total}), but also costs in the scenarios which are already cheap, as the system design does not need to compromise to keep costs down in the expensive cases, hence less generation and storage capacity can be used, reducing asset costs.}

\subsection{Value of building load monitoring} \label{sec:voi-results}

%% Overall
% Scatter plot prior and posterior designs on axes of total battery and pv capacities, with grid cap as colour
% Plot distributions of prior and posteriors costs?
% Might also be interesting to look at things like average grid cap for posterior solns
% Pick out average carbon emissions savings - interesting point is even if measurement only breaks even may still want to do it for the carbon

% Discuss interpretation and importance of VoI result - value is significant, but not huge

The On-Policy VoI framework presented in Sec. \ref{sec:on-policy} was used to quantify the economic benefit of uncertainty reduction from building load monitoring to support the design decision of sizing solar-battery systems and grid connection capacity for the district. A set of 256 hypothetical monitoring measurements were sampled using the probability models defined in Sec. \ref{sec:prob}, and the procedure of using Stochastic Programming to design (size) the system, and simulations to evaluate its performance, was repeated for each hypothesised measurement using samples drawn from the corresponding posterior distribution. This process considers the following questions;
\begin{itemize}
    \item[] ``Were the building monitoring systems to have returned a given measurement, reducing uncertainty in the building load profiles: What would the optimal district energy system design be? And what would be its average operational cost under the remaining load uncertainty?''
\end{itemize}

\begin{figure}
    \centering
    \begin{minipage}{.475\textwidth}
        \centering
        \includegraphics[width=\linewidth]{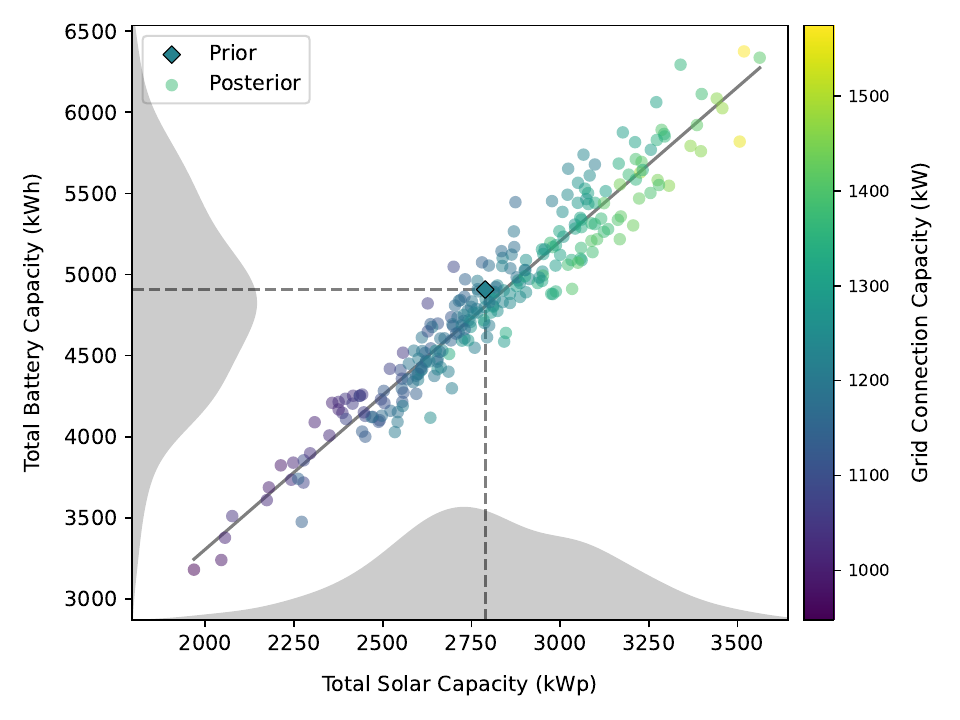}
        \caption{Total asset capacities in optimized system designs for each hypothesised load monitoring measurement.\\
        \textit{Prior system design indicated with diamond.}}
        \label{fig:posterior-designs}
    \end{minipage}%
    \hfill
    \begin{minipage}{.475\textwidth} % Could plot prior design on this chart, would it be useful?
        \centering
        \includegraphics[width=\linewidth]{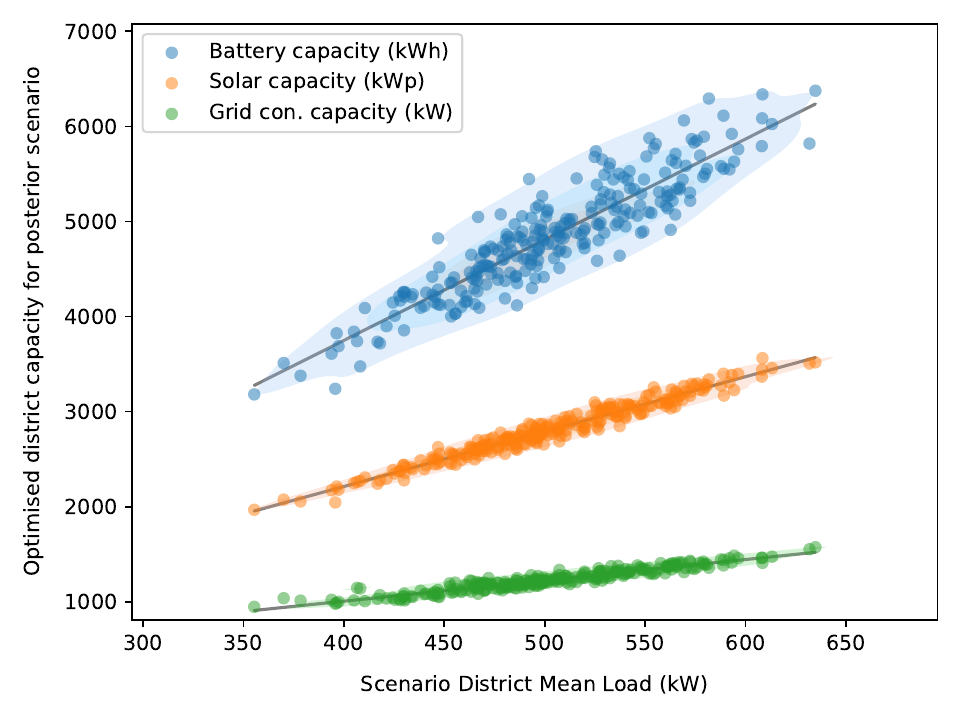}
        \caption{Correlation between total asset capacities in posterior designs and mean load over posterior sampled building load scenarios used for design (from scenario reduction).}
    \label{fig:posterior-designs-correlation}
    \end{minipage}%
\end{figure}

Fig. \ref{fig:posterior-designs} plots the total asset capacities of the optimized system design for each hypothesised load monitoring measurement (the posterior designs, each indicated with a circle), and compares them to the prior system design (indicated by a diamond).
It shows that the optimal asset sizes vary significantly with the load monitoring measurements, which provide information about the true energy usage behaviour of the district buildings, by approximately $\pm$20\% for each of solar, battery, and grid total capacities. These posterior optimal asset capacities are strongly correlated with the mean load of the posterior distribution, see Fig. \ref{fig:posterior-designs-correlation}.
% Comment on linear trends and relative spreads about these? (In appendix? Def. for thesis, discuss causality)

A sensitivity analysis studying the influence of input uncertainties on the optimal system design, such as those performed in \cite{mavromatidis2018UncertaintyGlobalSensitivity} \& \cite{pickering2019DistrictEnergySystem}, would stop at this point and conclude that the uncertainty reduction from building monitoring has a substantial impact on the optimal system design, and is therefore important information to gather for achieving a design close to the optimal for the district as built. However, this approach fails to address the actual goal of decision making. From the perspective of the decision-maker, they are not concerned with what decision is taken, but instead with how well on average that decision performs, i.e. its expected benefit/utility. Rather than the difference between the posterior designs and prior design themselves, the metric that is important for determining whether the building monitoring information is valuable for supporting decision making is the difference between their \textit{performance}. I.e. how much lower the lifetime cost of the district energy system is on average as a result of performing the system design with the building monitoring information, and the resulting reduced uncertainty. This metric is called the Value of Information (VoI), defined in Sec. \ref{sec:voi}.

\begin{figure}[t]
    \begin{minipage}{\textwidth}
        \subfloat[Total system costs.]{
            \includegraphics[width=0.475\linewidth]{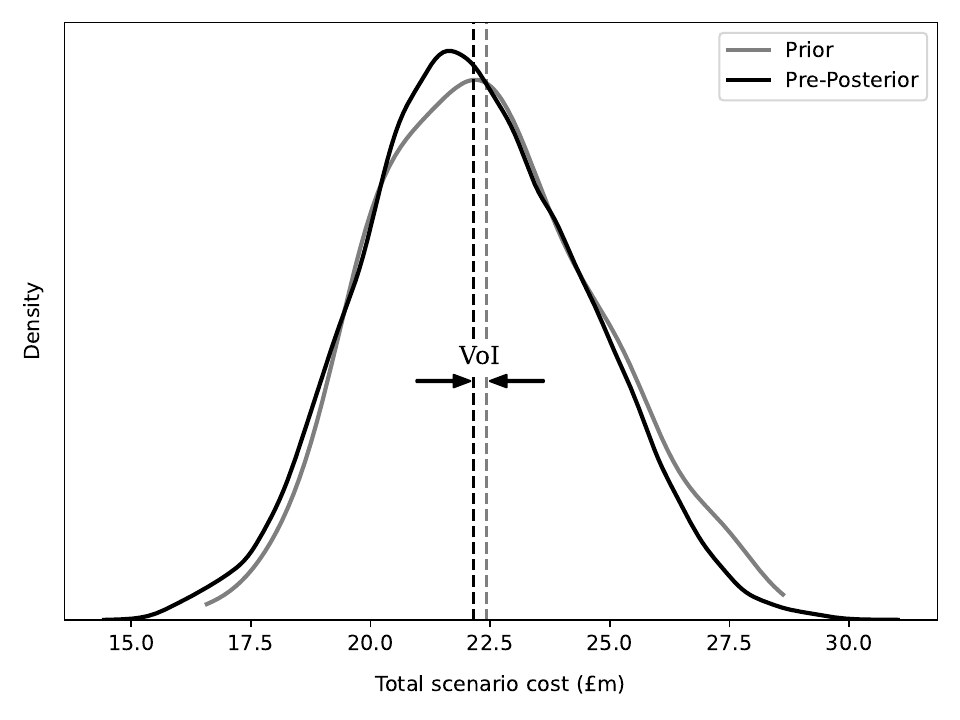}
        }%
        \hfill
        \subfloat[LCOEs.]{
            \includegraphics[width=0.475\linewidth]{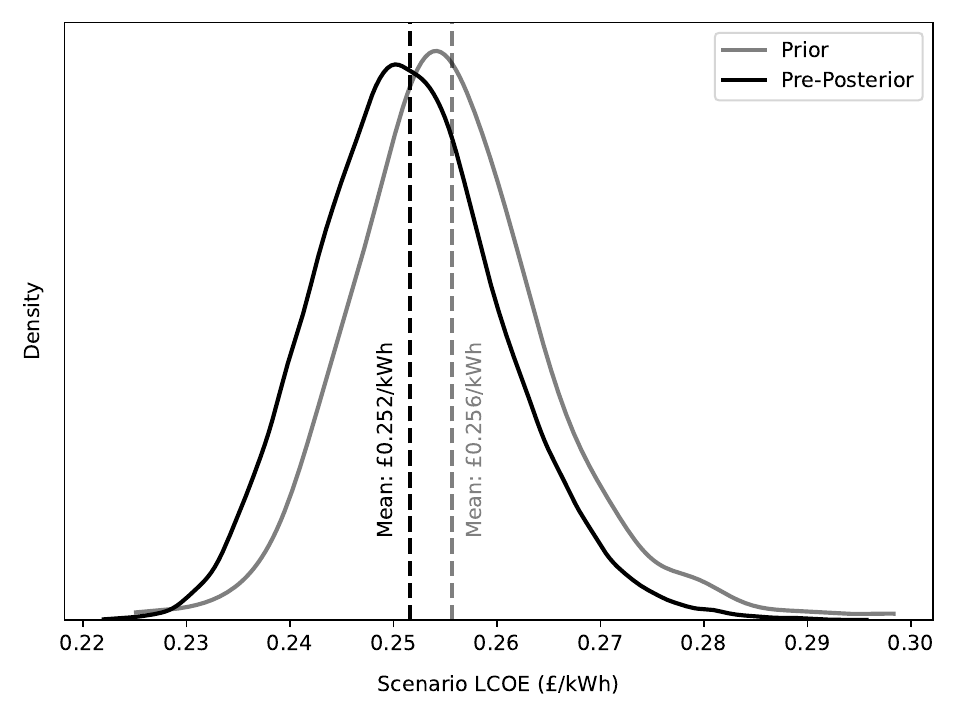}
        }
        \vspace*{-0.2cm}
        \caption{Comparison of simulated total system cost and LCOE distributions for prior and pre-posterior cases.} \label{fig:posterior-costs}
    \end{minipage}
\end{figure}

Fig. \ref{fig:posterior-costs} plots the distributions of total system costs and LCOEs achieved by the prior and posterior designs across the simulated building load scenario samples. The average lifetime cost for the posterior designs (the pre-posterior expected utility, Eqn. \ref{eq:OP-VOI-pre-posterior-utility}) is found to be £22.147m over the 20 year lifetime (£1.107m/year). Comparing to the prior expected utility of £22.432m, the Value of Information for building load monitoring is estimated at £284k, or 1.27\% of the prior cost.

This VoI result demonstrates that using building monitoring to support energy system design for this district reduces the average total cost of the energy system by under 1.5\%. This methodology leads to a conclusion that is very different from those provided by existing analyses, which suggest that the uncertainty reduction would have a large impact on the design task. Further, the insights provided by VoI are far more aligned with the goals of the decision maker.
% To add - whilst optimal designs vary significantly from prior, their performance does not

Gathering building monitoring data to improve the sizing of the solar-battery systems would only be worthwhile if the cost of collecting that data is less than the VoI value computed.
Installing the solar-battery systems was found in Sec. \ref{sec:prior-design} to reduce operational costs by on average £717k/year. Therefore, by itself the option cost of delaying installation of the solar-battery systems for one year to monitor the buildings and install an optimized system outweighs the benefits of that design optimization (provided by the uncertainty reduction from measurement). The actual cost of performing building monitoring would be higher due to costs associated with delaying the construction project. Hence, the VoI analysis shows that, for this particular energy system design task, reducing uncertainty in the building load profiles through monitoring is not economically worthwhile. I.e. the benefit it provides to improving decision making is less than its cost.
% Add more discussion of interpretation and pratical importance? Or leave for \ref{sec:practical-importance}?

\subsubsection{Relative importance of reduced uncertainties} \label{sec:contributions}

% Show results for VoI for partial information from each parameter separately and compare to overall
% Mean variation accounts for all of total VoI! This is important because this wouldn't be picked up by traditional experimental design (would suggest profile variance reduction), so is a value add from VoI (meaningful insight only it can provide)

The relative importance of each aspect of building load uncertainty (\texttt{type}, \texttt{peak}, \texttt{mean}) for supporting decision making was investigated by repeating the VoI analysis for uncertainty reduction in each parameter independently. Table \ref{tab:VOI-contributions} presents the results of this analysis. Similarly to the classical sensitivity analysis performed in Sec. \ref{sec:prior-design}, reducing the uncertainty in mean load is found to be of greater importance than uncertainty in the profile shape and peak load, providing roughly half the available benefit to decision making.
%However, the VoI based importance analysis provides two key insights which could not be identified by purely statistical analysis.

\begin{table}[h] % For thesis: add design figure for each case and discuss differences
    \centering
    \renewcommand{\arraystretch}{1.1}
    \begin{tabular}{c|c|c} \toprule \toprule
        Reduced uncertainties & Expected total cost (£m) [\%] & VoI (£k) [\%] \\ \midrule
        None (prior) & 22.432 [100.0\%] & -- \\
        \texttt{type} & 22.573\, [100.6\%] & 0\textsuperscript{\textdagger}\, [0.0\%] \\
        \texttt{peak} & 22.401\, [99.87\%] & 30\,\:\, [0.14\%] \\
        \texttt{mean} & 22.276\, [99.31\%] & 156\, [0.69\%] \\
        All & 22.147\, [98.73\%] & 284\, [1.27\%] \\
        \bottomrule \bottomrule
    \end{tabular}
    \caption{Expected system operating cost and VoI for separate uncertainty reduction in load parameters.}
    \label{tab:VOI-contributions}
\end{table}

\customfootnotetext{\textdagger}{This value is actually estimated to be -£141k (-0.63\%), but is clipped to 0 as VoI values are defined to be non-negative. This highlights the numerical error of the VoI calculations. However, even plus/minus this level of error, the VoI values are still small relative to the cost of measurement, and the conclusions remain valid.}

While uncertainty in the peak load and load profile shape of the buildings is responsible for a significant portion of the overall cost variability, collecting information which reduces uncertainty in only one of these parameters provides very little improvement for solar-battery system design. In fact, reducing uncertainty in the \texttt{type} parameter, which corresponds to obtaining a probabilistic model of the patterns and timing of energy use specific to each building, does not improve the designs at all. So, by itself, performing detailed occupancy modelling to generate precise time-of-use energy profiles for the buildings would not be helpful for supporting the design of this district energy system.

However, when all three uncertainties are reduce simultaneously, the VoI is significantly greater than the sum of the separate contributions\footnotemark. Therefore, if more precise information is available about the mean load of the buildings, then information about the peak load and load profile shape can be used to further improve the design. This result demonstrates that designers must consider not only whether they have the correct information to support their decision making, but the correct combination of information to unlock maximum value. Insights of this kind can only be obtained by applying the VoI methodology.

\footnotetext{This is possible as the Value of Information is not additive. Different measurements can either contain complementary information which is more useful for decision making when combined, or conversely can contain the same information, meaning the combination of measurements provides less total benefit \cite{difrancesco2023SystemEffectsIdentifying}.}

\newpage
\subsubsection{Alternative system scenarios} \label{sec:alt-systems}

% Discuss practical constraints on solar and show VoI is zero: decision devolves as battery set by solar (used for improving self-consumption mostly) and so when solar is known (i.e. best to be max) then there is little decision left to make on the battery
% Discuss setup with only 1 building: VoI higher as the relative variance in overall (district) mean load is higher (due to lack of CLT effect), discuss relevance for designing larger systems (the more buildings in scope the less useful info as mean values are most consistent - lower variance)

The same energy system sizing task was investigated in two alternative building system scenarios to study how the VoI of building load monitoring varied, and whether the conclusions regarding its economic usefulness changed from the base case. These two cases involve designing for a single building energy system, and designing for a district with constrained solar capacity.

Often adjacent buildings are owned by independent parties, and their energy systems are installed at different times with limited cooperation. So, solar-battery systems are frequently designed independently for each building, and for large consumers (e.g. industrial sites) each must contract their own grid connection capacity. Repeating the VoI analysis with a single building, i.e. without district effects, the VoI was found to be £99k, or 2.17\% of the prior cost of £4.568m. Whilst in this case the VoI has increased relative to the prior cost, it is still lower than the annual benefit of installing the solar-battery system, which is £145k/year. Therefore, it is still not economical to perform building monitoring to support design. It is proposed that the reduction in proportional VoI as more buildings are included in the district is caused by the variability (standard deviation) of the district mean load. Reducing uncertainty in mean load was found to be the most influential factor for improving system design in Sec. \ref{sec:contributions}, and the district mean load was the greatest driver of cost uncertainty, see Sec. \ref{sec:prior-design}. As mean load distributions are taken to be independent for each building, by the Central Limit Theorem, the standard deviation of the district mean load as a fraction of that load decays with the number of buildings in the district as $1/\sqrt{n}$. Therefore, measurement of the buildings provides less information on the district mean load the more buildings there are. Hence, the less valuable that information is for supporting design. This implies the more buildings there are in the district, the better decisions regarding total storage capacity for bulk energy arbitrage average out across the district, and the less sensitive they are to uncertainty in the mean load.

In many European countries, due to land constraints, buildings are built tall to accommodate more floor area, and do not have spare open spaces nearby. This limits their available roof area, and so the quantity of local solar generation they can accommodate to decarbonize their energy usage. The VoI analysis was repeated for a district of 5 buildings with a constraint on the maximum solar capacity (kWp) per building. This constraint was derived using roof area data from the Cambridge dataset \cite{langtry2024CambridgeUniversityEstates}. University buildings in Cambridge with around 100 kW mean load ($\pm$20\%) were found to have a roof area of approximately 1000 m$^2$. Assuming a solar panel power density of 0.15 kWp/m$^2$ \cite{gawley2022InvestigatingSuitabilityGIS,gagnon2018EstimatingRooftopSolar}, leads to an optimistic constraint of 150 kWp of solar capacity per building. With this constraint, the VoI of building load monitoring was found to be only £13k, or 0.05\% of the prior cost of £29.075m. Under the formulated operational objective, the primary benefit of the battery system is derived from arbitraging the cheap, zero-carbon solar energy generated by the building solar PV. This is why in the base case, the installed battery capacity of the optimal designs scale linearly with the solar capacity, see Fig. \ref{fig:posterior-designs}. For all scenarios in the solar constrained case, the maximum quantity of solar PV is installed, and so the value that can be derived from arbitraging that energy is fixed. Whilst the optimized system designs for each hypothesised measurement in the constrained case show some variability in the battery and grid capacities installed, these additional capacities provide only small additional benefit from arbitraging cheap and/or low carbon grid electricity, and these benefits are not much more than their costs. As a result, updating the system designs (battery sizings) with improved information provides very little benefit in this case.

\subsection{Planning risk reduction}

% Discuss standard deviation & range of outcomes pre and post measurement (highlight reduction) and discuss importance for project planning, even if there is little influence on mean cost; caveat of VoI framework is assumption of risk neutrality, which energy system designers certainly are not

In the VoI framework, decision makers are assumed to be risk neutral, and the benefit of data collection is quantified in terms of how much it is able to improve the \textit{average} performance/utility of decisions made. However, in the energy field, decisions makers are sensitive to risk, as it affects their ability to raise capital as well as the interest rates they can access, which impacts the cost of energy decarbonization and so their ability to afford it.
For the prior optimal design in the base case, the range of system lifetime costs over the simulations was found to be £16.580m to £28.616m, from a mean of £22.432m (see Fig. \ref{fig:prior-costs-total} in Sec. \ref{sec:prior-design}). This 54\% cost range could be problematic for the financial planning of the district energy system, particularly as the operational costs of energy system (electricity, carbon, and grid excess costs) vary from £6.360m to £18.396m.

\begin{figure}
    \centering
    \includegraphics[width=0.65\linewidth]{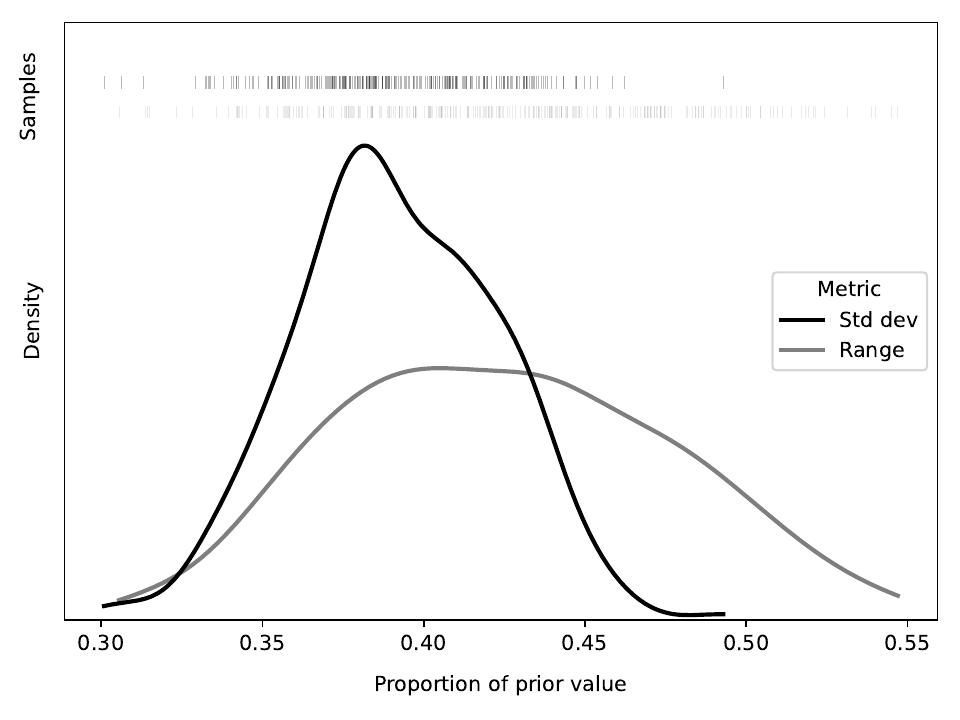}
    \caption{Distribution over hypothesised measurements of standard deviation and range of total cost distribution of optimized system design for each hypothesised measurement (simulated for scenarios drawn from corresponding posterior distribution).}
    \label{fig:posterior-variability}
\end{figure}

The information provided by load monitoring can be used to support this financial planning, as it reduces uncertainty in the operational cost of energy system. In the experiments, after a hypothetical monitoring measurement was taken, the system design was optimized for the corresponding posterior load distribution, and the operation of this system was simulated for scenarios sampled from that posterior, leading to a distribution of total system costs for the design. Fig. \ref{fig:posterior-variability} plots the distributions over the hypothesised measurements of the standard deviations and ranges of the simulated total system costs for each design (corresponding to a hypothesised measurement), as a proportion of the value for the total cost distribution for the prior design (shown in Fig. \ref{fig:prior-costs-total}). The standard deviation and range of the costs are significantly reduced after measurement, demonstrating that whilst designing the energy system with load monitoring information does not significantly reduce the average system cost, at the time of design and so financing, it substantially reduces the uncertainty in what the system cost will be. Therefore, there may be some additional value to this information for supporting financial decision making. This motivates the use of VoI to study the benefit of data collection to support risk averse decision making, which has yet to be done in an engineering context. This would require the expansion of utility/objective functions to account for the cost of risk. Existing studies have incorporated risk measures into district energy system design objectives to account for risk aversion \cite{pickering2019PracticalOptimisationDistrict}, and these methods are applicable to studying VoI.

\newpage
\subsection{Practical importance \& limitations of study} \label{sec:practical-importance}

The Value of Information calculations performed in this study (Sec. \ref{sec:voi-results}) demonstrate that, for the district energy system design task considered, hourly building load data is not worth collecting to support design, i.e. its benefit is less than its cost. This allows practitioners to avoid wasting time and resources on monitoring, and confirms that an opportunity for reducing the cost of decarbonizing building energy is not being missed. Further, showing that limited value is derived from reducing building load profile uncertainty indicates that population level priors constructed from monitoring data from existing buildings are sufficient for designing energy systems in unseen buildings. This provides the first numerical evidence to suggest that current industry practice of designing energy systems using standard building profiles achieves near optimal design performance. However, it is strongly recommended that using Stochastic Optimization techniques which seek designs that perform well over the distribution of building behaviours is critical.

Sec. \ref{sec:contributions} found that reducing uncertainty in only the mean load provided most of the available decision support benefit. This result is highly pertinent for the design of both new building energy systems, and energy retrofits of existing buildings. For new buildings, while in this case study load monitoring is not worthwhile to support design, the reduction in mean load uncertainty does still provide significant benefit. Therefore, effort should be directed to producing an accurate statistical estimate of building mean load, for instance through detailed building usage modelling, i.e. at room level. Communication between the energy system designer and building owners/occupants regarding the intended use of the building is critical to enabling accurate load estimates, and so low cost energy system designs. For the case of building retrofits, monthly gas and electricity metering data are available. This enables a very good estimate of the mean building electrical load after retrofit, and so using hourly metering data for system sizing would provide little additional benefit.

In this study, uncertainty in building load and solar generation only were considered for the sake of clarity when demonstrating the VoI methodology. However, several other important factors in the district energy system are subject to significant uncertainty, such as electricity prices (which are determined by the future grid generation mix and weather conditions), asset operational lifetimes which impact their effective cost, and battery degradation (neglected in this model) which affects the efficiency and capacity of the battery units over their lifetime. More detailed system design work should consider these uncertainties in the Stochastic Optimization model to include their influence on the design task and resulting cost distributions. A key difference with these uncertainties is that they are not practically measurable at the time of design, and so should not be studied using VoI, as the resulting values would not be meaningful.

Performing VoI calculations is highly computationally intensive. In the On-Policy VoI framework, system designs must be both determined and evaluated for a large number of hypothesised measurements to achieve sufficient statistical accuracy in the mean utility estimates to enable the correct identification of the true VoI value over statistical error. This limits the complexity/resolution of building energy system models that can be studied with the framework. However this computational cost is minimal compared to the potential cost savings of avoiding project delays to gather unnecessary building monitoring data. The computational requirements of the VoI calculations from this study are discussed further in \ref{app:numerics}.

Finally, the key critique of the VoI framework is that VoI values computed are only valid under the particular model formulation studied, including its assumptions \cite{langtry2024RationalisingDataCollection}. VoI results cannot be generalized, and changes in the problem setup can significantly alter the VoI, as seen when introducing solar capacity constraints in Sec. \ref{sec:alt-systems}. Therefore, further investigation of VoI for supporting building energy system design is required to confirm whether the conclusions made in this study are consistent across different building energy systems (e.g. those in different climates or with different energy usage patterns), design objectives, and modelling assumptions. The VoI methodology should also be used to investigate the importance of reducing other uncertainties in buildings (such as thermal properties of the envelope), and whether performing other types of measurements could provide significant improvements for the design of energy systems.

%% file: sections/conclusions.tex
\section{Conclusions} \label{sec:conclusions}

% Summarise results and importance.
% Discuss further work: application to other energy system design \& control tasks (on-policy voi enables study of practical control); study of other reducable uncertainties; further study of VoI under risk aversion; study of load profile distributions across multiple building types?

This study investigated the economic benefit of gathering hourly building energy usage data to support the design of a district energy system, specifically involving the sizing of distributed solar-battery systems to decarbonize building energy usage under grid constraints. A case study district energy system containing 5 university buildings was considered, with probabilistic models of the building load profiles constructed using historic electricity metering data from the Cambridge University estate.

The uncertainty in building load was found to have a significant impact on the total system operating cost, causing a variation of $\pm$30\% for the energy system designed without measurement information available. Additionally, reduction in building load uncertainty from monitoring had a significant effect on the optimized energy system design, with total installed capacities of solar, battery, and grid varying by roughly $\pm$20\% across the hypothesised measurement values. Therefore, existing methods which stop at either of these two points would conclude that using hourly metering data would have a significant effect on system design.

However, performing a Value of Information analysis using the On-Policy VoI framework extension proposed in this work, demonstrated that reducing load uncertainty via monitoring reduces the overall energy system cost by under 1.5\% on average through improved design. This was found to be less than the cost savings achieved by installing solar-battery systems to reduce operating costs over the period needed to monitor the buildings and install systems with improved design. Hence, collecting hourly building monitoring data was determined to be not economically worthwhile for supporting the design of the studied district energy system.
Additionally, reducing uncertainty in only the mean load was found to provide most of the available decision support benefit. Therefore for building retrofits where the mean building load can be well estimated from monthly usage data, using hourly metering data provides little additional benefit for system design.
This demonstrates the need for Value of Information analysis to justify and prioritize expenditure on data collection to support decision making, as current analyses can lead to erroneous conclusions regarding the benefit of uncertainty reduction.

These results are important for practitioners, as identifying the low value of load uncertainty reduction for supporting district energy system design allows wasteful delays to monitor building loads\footnotemark to be avoided. Additionally, it indicates that population level priors constructed using monitoring data from existing buildings are sufficient for energy system design. This provides the first numerical evidence to support the sufficiency of existing industry practice of using (distributions of) standard building load profiles for energy system design. % Add comment about need to assess whether deterministic profiles are any good?
\footnotetext{Note that monitoring systems would still be installed in the buildings, as real-time building load data is critical for effectively controlling the solar-battery systems. This work quantifies the value of having access to this data \textit{before} the solar-battery systems are designed, to reduce uncertainty in the building loads and improve sizing decisions. The results demonstrate that it is not worthwhile delaying the installation of the solar-battery systems to collect data from the monitoring systems, as the benefit to decision making is less than the cost of doing so. Over time, operational data from the monitoring systems can be used to improve the building load distributions which are then used for future designs.}

However, as the VoI framework provides no generalization guarantees, this study can only indicate that collecting load monitoring data is not worthwhile for supporting the design of the particular case study district investigated. Further research is required to determine whether the benefit of load monitoring remains below its cost across different energy system setups, different design objectives, and with the inclusion of additional uncertainties. Additionally, further work is required to determine whether uncertainty reduction can provide significant benefit to supporting financial planning through risk reduction.

Value of Information analysis provides a clear methodology for addressing the research gap of quantifying the benefits of uncertainty reduction for supporting decision making in building energy systems. Further, the On-Policy VoI framework proposed in this work enables the study of VoI in the context of more complex decision making tasks, such as energy system control using Model Predictive Control or Reinforcement Learning. With the ability to quantify and compare the benefit of data collection strategies, energy system designers and operators will be able to reduce the overall cost of decarbonizing energy usage, through either avoiding wasteful expenditure on low insight measurement, or identifying new measurement opportunities which enable the greatest cost reductions.

%% file: sections/endmatter.tex
\section*{CRediT authorship contribution statement}

\textbf{Max Langtry}: Conceptualization, Software, Methodology, Investigation, Writing - Original Draft, Writing - Review \& Editing
\textbf{Ruchi Choudhary}: Supervision, Writing - Review \& Editing

\section*{Declaration of competing interests}

The authors declare that they have no known competing financial interests or personal relationships that could have appeared to influence the work reported in this paper.

\section*{Data availability}

All code and data used to perform the experiments in this study are available at \url{https://github.com/mal84emma/Building-Design-VoI}. The numerical results presented are available at \url{https://zenodo.org/records/15614750}.

\section*{Acknowledgements}

Max Langtry is supported by the Engineering and Physical Sciences Research Council, through the CDT in Future Infrastructure and Built Environment: Resilience in a Changing World, Grant [EP/S02302X/1].

This work was performed using resources provided by the Cambridge Service for Data Driven Discovery (CSD3) operated by the University of Cambridge Research Computing Service (\href{https://www.csd3.cam.ac.uk/}{www.csd3.cam.ac.uk}), provided by Dell EMC and Intel using Tier-2 funding from the Engineering and Physical Sciences Research Council, Grant [EP/T022159/1], and DiRAC funding from the Science and Technology Facilities Council (\href{https://dirac.ac.uk/}{www.dirac.ac.uk}).

%% file: sections/appendices.tex
\section{Case study system parameter specification} \label{app:case-study-params}

Common parameter values specifying the district energy system, probabilistic model of building load, and procedure for sampling from that model, are used across all experiments. Table \ref{tab:system-params} provides values for the parameters of the district energy system model, including costs defining the objective/utility, simulation durations, and assumptions regarding battery performance and grid control. All parameters for the probabilistic model of building load are specified within the definition of the probabilistic model in Sec. \ref{sec:prob}. Table \ref{tab:sampling-params} provides the settings used when sampling for the defined distributions, and performing scenario reduction using the Fast Forward technique \cite{heitsch2003ScenarioReductionAlgorithms}, as implemented in \cite{gioia2023ScenarioReducer}.\\

% system parameters
\begin{table}[h]
    \centering
    \renewcommand{\arraystretch}{1}
    \renewcommand\cellset{\renewcommand\arraystretch{0.33}%
        \setlength\extrarowheight{0pt}}
    \begin{tabularx}{\linewidth}{lccX} \toprule \toprule
        \multicolumn{1}{>{\centering\arraybackslash}c}{Parameter} & Units & Value & \multicolumn{1}{>{\centering\arraybackslash}c}{Note/Refs} \\
        \midrule \midrule
        Carbon cost & £/kgCO$_2$ & 1 & \makecell[tl]{Chosen to be much greater than current carbon\\trading prices, c. £0.1/kgCO$_2$ \cite{desnz2022UKETSCarbon,bloomberg2024EUETSMarket}, to reflect\\ building owners' decarbonization priorities.} \\
        Battery capacity cost & £/kWh & 750 & \cite{mottmacdonald2018StorageCostTechnical,forbeshome2024HowMuchDoes} \\
        Solar capacity cost & £/kWp & 1500 & \makecell[tl]{\cite{gawley2022InvestigatingSuitabilityGIS,e.onenergy2024SolarPanelCost}; corresponds to LCOE of £58.6/MWh,\\ in line with expected value \cite{desnz2023ElectricityGenerationCosts}} \\
        Grid connection capacity cost & £/kW/day & 0.263 & \makecell[tl]{Chosen to be much greater than current costs,\\ c. £0.05/kVA/day \cite{easternpowernetworks2023EasternPowerNetworks}, to reflect future grid\\constraints. Converted to per kW value assuming\\a power factor of 0.95 \cite{eepower2022PowerFactorDetermining,anderson2021PowerFactorReactive}.}\\
        Grid connection excess charge & £/kW/day & 1.053 & As above. Current value c. £0.08/kVA/day \cite{easternpowernetworks2023EasternPowerNetworks}. \\
        Simulation duration, $T$ & Hours & 8760 & \\
        System lifetime & Years & 20 & \\
        Design grid safety factor & -- & 1.25 & Determined from initial experiments. \\
        Operation grid safety factor & -- & 1.01 & Determined from initial experiments. \\
        Battery discharge ratio, $\delta$ & kW/kWh & 0.4 & \makecell[tl]{Value for Tesla Powerwall \cite{forbeshome2024HowMuchDoes}} \\
        Initial state-of-charge, $\textrm{SoC}^0$ & -- & 0 \\
        \bottomrule \bottomrule
    \end{tabularx}
    \caption{Parameter values for district energy system model used in experiments.}
    \label{tab:system-params}
\end{table}

% sampling parameters
\begin{table}[h]
    \centering
    \renewcommand{\arraystretch}{1}
    \begin{tabular}{lc} \toprule \toprule
        \multicolumn{1}{>{\centering\arraybackslash}c}{Parameter} & Value \\
        \midrule \midrule
        No. samples from prior distribution & 1000 \\
        No. samples from each posterior distribution & 256 \\
        MCMC sampling burn-in period & 256 \\
        MCMC sampling thinning factor & 10 \\
        No. of reduced scenarios used in Stochastic Program & 10 \\
        \bottomrule \bottomrule
    \end{tabular}
    \caption{Settings for sampling from probabilistic load model used in experiments.}
    \label{tab:sampling-params}
\end{table}

\newpage

\section{Details of prior system design} \label{app:prior-design}

% Include details of prior system design and behaviour
% - building breakdown of capacities; discuss how small number of scenarios leads to variation, would expect uniformity as n_samples grows large as all buildings have same prior
% - provide breakdown of costs
% - figures of optimized operation; link to interactive online version
% - discuss features of operation behaviour
% - discussion of factors dominating design, make suggestions about causality from interpreting operation behaviour (when solar unconstrained, grid capacity is set by solar peaks, batteries are needed to shave peak generation, batteries are sized for arbitrage not for power, etc.) - reader can explore further using interactive version of plot available online at ...

The sizing of grid connection capacity and solar-battery systems for each building determined by the prior design optimization is presented in Table \ref{tab:prior-design}. As all buildings in the district are taken to have identical probabilistic load models and solar generation potentials, in the perfect case, the design optimization would allocated equal asset capacities in all buildings, as on average they behave identically. However, Table \ref{tab:prior-design} shows significant variation in the solar-battery sizings across the buildings. This occurs as only a small number of reduced scenarios (10) are included in the Stochastic Program model, due to the cubic computational complexity of Linear Programming limiting the number of scenarios which can be considered. Random variation in the mean building loads across the reduced scenarios results in non-uniform sizing across the district.

Table \ref{tab:prior-costs-breakdown} provides a breakdown of the system operating cost contributions for the prior design across the 256 simulated building load scenarios.\\

\begin{table}[h]
    \centering
    \renewcommand{\arraystretch}{0.8}
    \begin{tabular}{c|ccc} \toprule \toprule
        Building no. & Solar capacity (kWp) & Battery capacity (kWh) & Grid con. capacity (kW) \\
        \midrule \midrule
        0 & 617 & 1,119 & -- \\
        1 & 538 & 973 & -- \\
        2 & 497 & 844 & -- \\
        3 & 583 & 983 & -- \\
        4 & 554 & 989 & -- \\ \midrule
        Total & 2,789 & 4,908 & 1,227 \\
        \bottomrule \bottomrule
    \end{tabular}
    \caption{Installed asset capacities for prior system design.}
    \label{tab:prior-design}
\end{table}

\begin{table}[h]
    \centering
    \renewcommand{\arraystretch}{0.8}
    \begin{tabular}{c|ccc} \toprule \toprule
        Cost & Mean (£m) & Percentage (\%) & Std. dev. (£m) \\
        \midrule \midrule
        Total & 22.432 & 100.0 & 2.282 \\
        \gray{LCOE (£/kWh)} & \gray{0.256} & -- & \gray{0.00971} \\
        \midrule
        Electricity & 7.416 & 33.1 & 1.357 \\
        Carbon & 4.795 & 21.4 & 0.934 \\
        Grid excess & 0.0 & 0.0 & 0.0 \\
        Grid connection & 2.356 & 10.5 & -- \\
        Battery & 3.681 & 16.4 & -- \\
        Solar & 4.183 & 18.6 & -- \\
        \bottomrule \bottomrule
    \end{tabular}
    \caption{Breakdown of cost contributions over sampled scenario simulations for prior system design.}
    \label{tab:prior-costs-breakdown}
\end{table}

\newpage

\begin{figure}[t]
    \begin{minipage}{\textwidth}
        \centering
        \subfloat[Summer.]{
            \includegraphics[width=\linewidth]{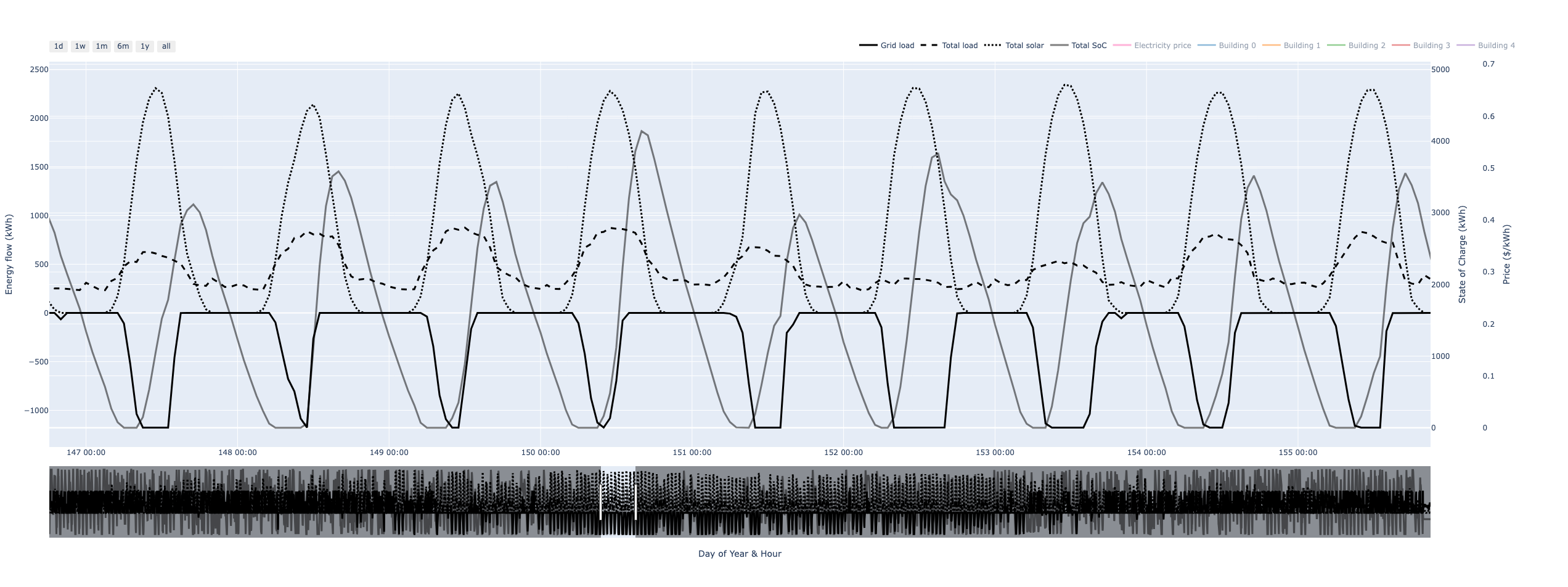}
            \label{fig:prior-simulation-summer}
        }

        \subfloat[Winter.]{
            \includegraphics[width=\linewidth]{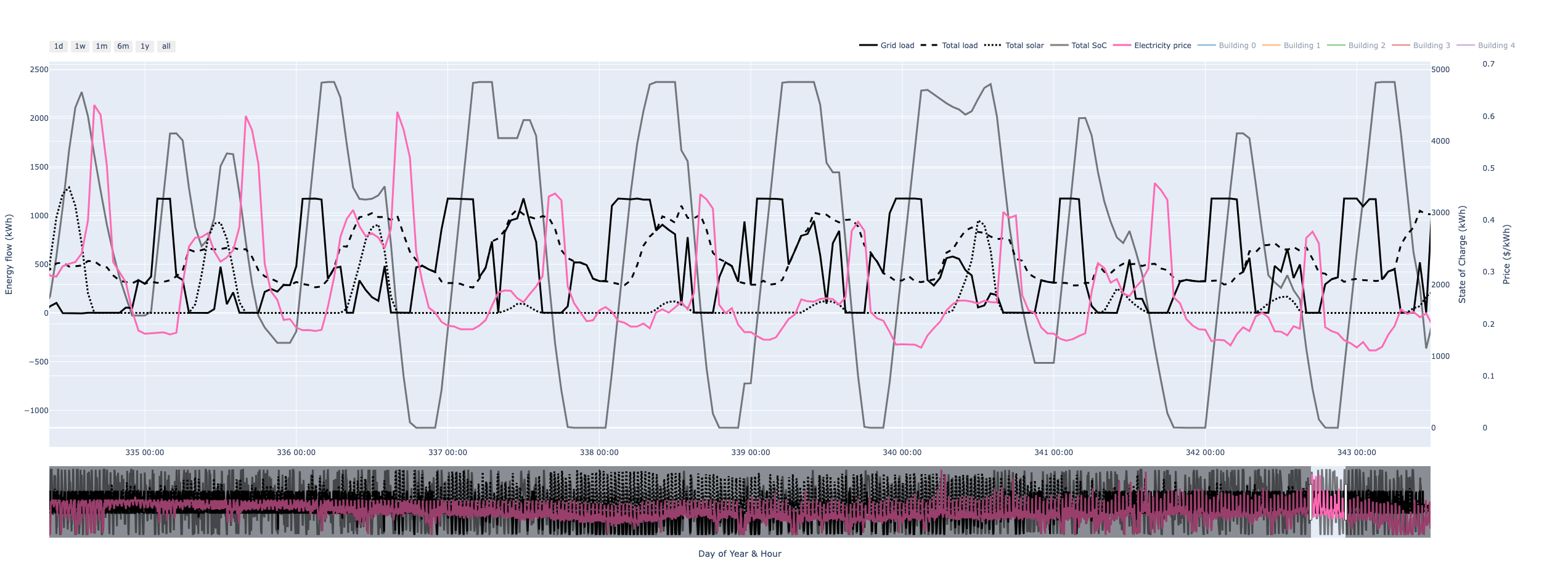}
            \label{fig:prior-simulation-winter}
        }
        \vspace*{-0.2cm}
        \caption{Illustration of operation of prior system across seasons.}
        \label{fig:prior-simulation}
        % add link to interactive example control plot
        \vspace{0.1cm}
        \small{\textit{Interactive version of plot available online} \href{https://mal84emma.github.io/Building-Design-VoI/example_prior_system_simulation.html}{\textbf{here}}.}
    \end{minipage}
\end{figure}

Observation of the operational behaviour of the district energy system provides insight into the factors driving system design. Fig. \ref{fig:prior-simulation} plots the overall loads and battery state-of-charge in the district when operating over periods in both the summer (Fig. \ref{fig:prior-simulation-summer}) and the winter (Fig. \ref{fig:prior-simulation-winter}) for one of the sampled load scenarios. Readers can investigate the operational behaviour of the district energy system further using an interactive version of plot available online at \url{https://mal84emma.github.io/Building-Design-VoI/example_prior_system_simulation.html}.

Comparing Figs \ref{fig:prior-simulation-summer} \& \ref{fig:prior-simulation-winter} shows that the net peak powers caused by local solar generation in the summer are significantly larger than those caused by building load in the winter. As a result, in the base case without constraints on the solar capacity, the required grid connection capacity is determined by the quantity of solar PV, and the optimization trades-off grid connection cost with additional cheap, zero-carbon solar energy. This implies that consideration of uncertainty in solar generation in the design optimization is important, as the capacity factor of solar generation will determine the scaling of this trade-off (noting that the peak power is well known as it is approximately the capacity in all years). This behaviour of the district energy system regularly exporting large quantities of energy onto the grid at around noon in the summer may cause issues for regulation of the power grid, as discussed in \cite{iweh2021DistributedGenerationRenewable}, and will require either adaptation of grid infrastructure to handle reverse power flow, or differentiated pricing for grid export capacity to discourage energy prosumers (customers with local generation) from flooding the grid with solar energy at these peak times. This may then increase the need for local storage to prevent surges in grid export to manage grid stability.

For the majority of the summer, the district draws no power from the grid, though it provides significant quantities of energy to the grid. Therefore, over the summer this district would have zero energy bills. Additionally, as in this model the energy system operator is not paid for supplying energy to the grid, the solar generation capacity is therefore being set by its ability to provide cheap, zero-carbon energy during the spring and autumn to offset grid electricity. It is therefore correlated with the district mean load, as this determines the quantity of energy which must be offset. In future energy systems with large quantities of solar generation, it is likely that during peak generation times the price of grid electricity would fall to approximately zero, or potentially negative, as instantaneous generation exceeds energy demand.

Full depth battery charge cycles are observed in the winter but not the summer. This indicates that the battery capacities are determined by the financial benefit of arbitraging cheap grid electricity for use by the buildings at peak times. These capacities are then sufficient for maximising the usage of local solar energy generation, and managing power peaks to avoid exceeding the grid connection capacity. Fig. \ref{fig:prior-simulation-winter} shows that the Model Predictive Controller successfully performs this price arbitrage of energy. On most days of the simulation, the power draw from the grid is near zero at peak demand times (around 5pm). In fact, this distributed generation and storage system reverses the timing of the peaks as the controller is not directed to smooth power usage.

As the district mean load determines the quantity of energy which must be arbitraged to reduce the cost of satisfying building loads at peak times, via the same mechanism as for solar capacity, the optimal battery capacity is therefore also determined by the district mean load. This justifies the correlation between optimized solar and battery capacities observed in Fig. \ref{fig:posterior-designs}.

\newpage

\section{Investigation of numerical accuracy \& computational requirements} \label{app:numerics}

\subsection{Statistical accuracy}
%% Show MC estimate convergence to verify statistical accuracy of expectations
% Estimate converge well within no. of samples, so sufficient samples have been used
% Improved sampling technqiues (MCMC variants) could be used to improve sample efficiency (no. of effective samples, e.g. from tails) and reduce computational cost
% Also, we could probably get away with fewer samples and still have a decent error
% Compare standard errors on prior & posteriors separately vs standard error on VoI (treating as a single mean); 2nd option seems much more reasonable given visual convergence

As the Value of Information is the difference of two expectations, the prior and pre-posterior expected utilities (see Eqn. \ref{eq:EVII}), which are estimated using Monte Carlo sampling, the computed VoI values are statistical estimates of some underlying true value. The statistical accuracy of these estimates must therefore be considered before firm conclusions can drawn regarding whether data collection is worthwhile.

\begin{figure}[b!] % bounds are P95 (1.96sd)
    \centering
    \includegraphics[width=0.7\linewidth]{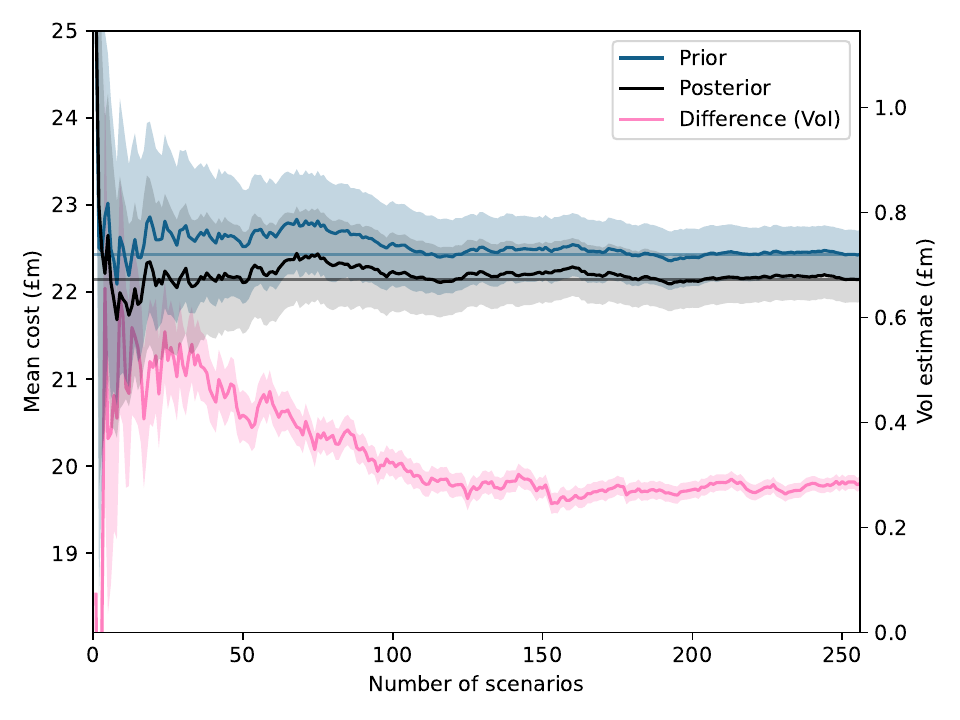}
    \caption{Convergence of Monte Carlo estimates of expected utilities with number of samples from building load distribution.}
    \label{fig:MC-convergence}
\end{figure}

Fig. \ref{fig:MC-convergence} plots the convergence of the prior and pre-posterior expected utility estimates, as well as the VoI estimate, as the number of samples from the building load distribution (scenarios, with corresponding designs and simulations) used increases. It also plots 95\% confidence intervals for the expectations using the standard error on the sample mean. For the VoI estimate, the VoI is considered as a single expectation, and the standard error is computed for the convergent sequence of VoI estimates. The expectation estimates are seen converge within the 256 samples used in the experiments, providing very similar estimates for all metrics after only 128 samples. This indicates that the VoI values computed in the experiments have sufficient statistical accuracy to support the hypothesis that the VoI is smaller than the cost of measurement, and therefore building monitoring is not financially worthwhile for supporting system design.

In this study a naive sampling scheme was used to sample from the building load distribution. The use of improved sampling techniques which provide greater sample efficiency for the load distribution would speed up the convergence of the expectation estimates, and so reduce the computational cost of achieving sufficient statistical accuracy in the VoI estimates. Further work is required to properly quantify the statistical error of VoI estimates, and develop techniques for determining the number of samples required to achieve sufficient statistical accuracy to support conclusions drawn from the VoI, i.e. whether it is greater or less than some cost threshold. Such developments are necessary to reduce the computational cost of VoI calculations, to allow the study of complex, practical engineering decision problems, whilst providing confidence that the data collection strategy recommendations they produce are correct with high probability.

\subsection{Cost estimation accuracy} \label{app:cost-accuracy}
%% Discuss accuracy of SP objectives - surprisingly good with only 10 scenarios and extremely simple scenario reduction (also the case for 1 building?) - but this couldn't be known a priori
% Briefly compare SP objectives to evals and comment on gap (very small gap considering no. of SP scenarios) - discuss why (little planning benefit from more than 48hrs of foresight as energy storage only for short durations)
% We claimed that the SP objective *may* not be a good approximation of the actual cost and could obscure the VoI - check this (for both prior & posterior SPs); initial checks seem to suggest objective is very good approximation for 2 reasons: 1 very little intra-day storage interaction (so perfect foresight provides little benefit), grid cap FoS well calibrated (very small grid excess charges in operation) => for this case could get rid of utility eval step and save lots of computation; this is actually quite impressive considering such basic scenario reduction is used as well as only 10 MC samples for objective
% Discuss practical importance for buiding design using SPs and the trustworthiness of their cost predictions - obviously limitation is that we used perfect foresight, and practical forecasts lead to worse performance, as demonstrated in Annex paper (cite)
% Removing simulation step would save c. 80% of runtime and make VoI calculations far more achievable, i.e. 100 LPs could be solved in parallel on HPC

The On-Policy VoI framework proposed in Sec. \ref{sec:on-policy} enables the VoI to be computed for complex decision problems where policies which do not necessarily provide an accurate estimate of the decision utility must be used for computational tractability. For the case study considered in this work, Stochastic Programming (SP) is needed to optimize the energy system sizing as the number of decision variables and simulation time makes the application of stochastic global optimization methods directly to the simulator intractable. However, the SP's estimate of mean system cost (expected utility) suffers from two sources of inaccuracy.

Firstly, the SP can only include 10 scenarios in its Monte Carlo estimate of mean operational cost within its objective function to run in a manageable computation time. This leads to potentially significant statistical error in the estimate of cost that is minimized. An attempt is made to manage this by using scenario reduction to select the 10 samples/scenarios that provide the best statistical representation of the population. However, as the space of possible load profiles is extremely large, only simple summary statistics could be used as the targeted statistical metric to represent, meaning the selection of scenarios to include in the SP does not have a direct link to the distribution of operational costs. Secondly, modelling error is introduced as the Linear Program formulation of the decision problem requires the assumption that the energy system is controlled with perfect foresight of the operational conditions, i.e. the weather and electricity price is known perfectly for all future periods during control. This assumption is not realistic and can lead to the SP model underestimating operational costs as the system is able to pre-empt difficult operational conditions and make unrealistically good control decisions in advance to reduce overall cost. In the case of managing the grid impact of energy usage under load uncertainty this assumption can be particularly over-optimistic, as unforeseen load conditions can lead to the battery state-of-charge not being sufficient to manage large peak loads, leading to substantial grid excess charges. Assuming perfect foresight allows the system to manage the battery level to avoid these instances, and causes a significant underestimate of operational cost. In fact, it is this over-optimism regarding the system's ability to manage peak loads that leads to the need to include a factor-of-safety on grid capacity during design.

Therefore, as these model and statistical errors could lead to significant inaccuracy in the SP's estimate of operating costs, a simulation step was included to provide a more accurate estimate of the true costs. These simulations used a Model Predictive Controller with only 48 hours of perfect foresight\footnotemark, and evaluated a large number of sampled load scenarios to overcome both the model and statistical inaccuracies.
\footnotetext{This controller is still unrealistic, as perfect foresight even over a limited planning horizon leads to practically unachievably low operational costs, as demonstrated in \cite{langtry2024ImpactDataForecasting}. Whilst it would not be feasible to train practical prediction algorithms to use in each simulation, a controller using synthetically noisy forecasts calibrated to the properties of practical forecasting methods could be used to improve the realism of the simulated operation and resulting costs.}

\begin{figure}[t]
    \centering
    \includegraphics[width=0.5\linewidth]{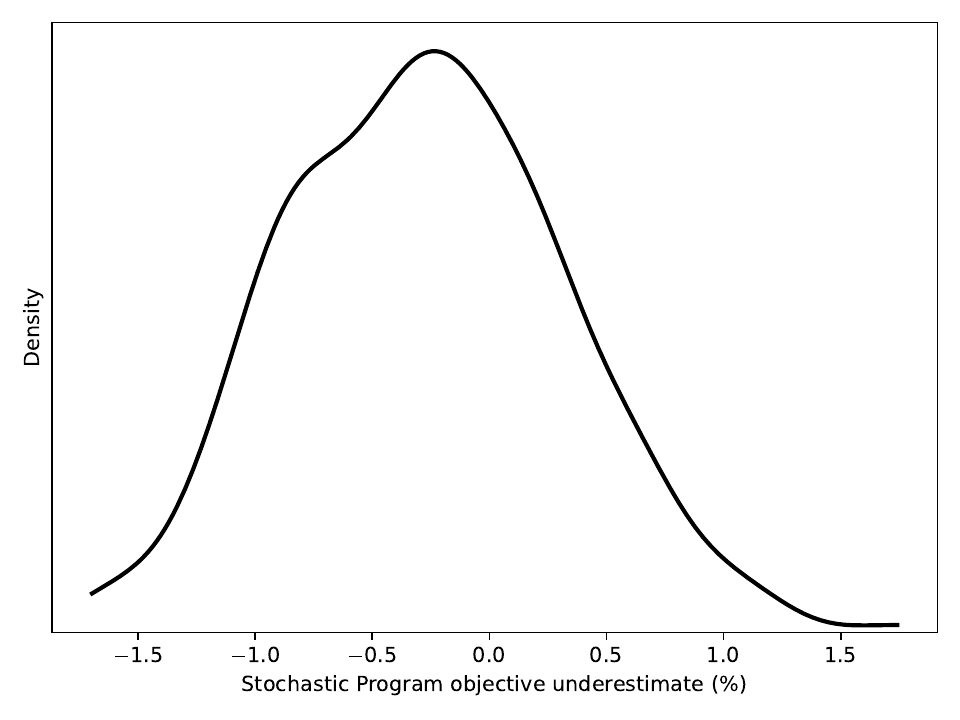}
    \caption{Distribution of Stochastic Program objective value errors compared to simulated mean operating costs for posterior designs in base case.}
    \label{fig:post-SP-obj-error}
\end{figure}

The accuracy of the Stochastic Program model objective was investigated by comparing it to the mean system operating costs determined by simulation. Table \ref{tab:SP-obj-error} provides both the mean error and mean absolute error of the SP objective for system designs performed with both the prior and posterior distributions, across the range of cases considered in the experiments. Fig. \ref{fig:post-SP-obj-error} plots the distribution of objective errors for  the SP designs performed for each measurement scenario in the base case, where all uncertainties were reduced.\\

\begin{table}[h]
    \centering
    \renewcommand{\arraystretch}{0.8}
    \begin{tabular}{c|cc} \toprule \toprule
        Case & Mean error (\%) & Mean absolute error (\%) \\
        \midrule \midrule
        \multicolumn{3}{>{\centering\arraybackslash}l}{\footnotesize \qquad Prior cases (as considered in Sec. \ref{sec:alt-systems}) - point estimates} \\[1ex]
        Base & -0.345 & 0.345 \\
        Solar constrained & -0.423 & 0.423 \\
        1 building & -3.062 & 3.062 \\
        \midrule
        \multicolumn{3}{>{\centering\arraybackslash}l}{\footnotesize \qquad Posterior cases (as defined in Table \ref{tab:VOI-contributions} of Sec. \ref{sec:contributions}) - av. over 256 msrmnts.} \\[1ex]
        \texttt{type} & 0.683 & 0.811 \\
        \texttt{peak} & -0.113 & 0.899 \\
        \texttt{mean} & -0.633 & 1.052 \\
        All (base) & -0.271 & 0.525 \\
        \bottomrule \bottomrule
    \end{tabular}
    \caption{Errors in SP estimates of mean system operating costs for optimized designs (under-estimation +ve).}
    \label{tab:SP-obj-error}
\end{table}

The results show that the Stochastic Program is able to estimate the mean operating cost within around 0.25-1\% (expect for the single building case where the error is significantly higher), and in fact more frequently over-estimates the cost than under-estimates. This result is somewhat surprising considering that only 10 scenarios are considered in the SP objective, and these scenarios were selected in a simplistic manner. From the plots of operational behaviour (Fig. \ref{fig:prior-simulation}) it can be seen that battery charge is rarely held for more than 72 hours, which indicates that during operation, assumption of complete perfect foresight would not lead to significant under-estimation of operational costs. However, this also demonstrates that the factor-of-safety used during design is performing well as the system does not end up in situations where is does not have the state-of-charge needed to manage peak loads. Additionally, as district mean load was one of the metrics used in scenario reduction, and was found to be the key driver of operational cost (Sec. \ref{sec:prior-design}), the scenario reduction was able to well represent the distribution of costs through correlation.

However, whilst the SP objective error is relatively low, it is approximately the same magnitude as the VoI, which was 1.27\% in the base case. An additional 1\% cost error added to the VoI estimate could have led to the conclusion that building monitoring was economically worthwhile, as it would have put it at around the same level as the estimated cost of performing the measurement. However, the impact of the SP objective error on the VoI estimate is complex, as it depends on the error in the difference between the prior and pre-posterior cost estimates, which is challenging to study. Ultimately, verifying that the error is sufficiently low may as computationally expensive as performing the simulations required by the On-Policy VoI calculation.

\subsection{Computational cost}

The computational cost of the experiments performed in this study was very large, with each VoI calculation taking roughly 10 days of computation time on a powerful desktop workstation\footnote{32 core, 2.90 GHz Intel Xeon Gold CPU with 256 GB of DDR4 RAM.} with a parallel processing workflow. Performing the simulations took approximately 80\% of runtime.

Reducing the number of samples required by the Monte Carlos estimates of expected utilities, or removing the need for simulation to accurately estimate utilities, would greatly decrease the computational cost of VoI calculations.
Given that the SP objectives are found to be reasonably accurate estimates of the simulator costs in \ref{app:cost-accuracy}, a good estimate of the VoI could be obtained in only 20\% of the original runtime.
A key advantage of VoI calculations is that they are highly parallelizable, as each scenario can be evaluated independently, making them well suited for distributed computation. I.e. the optimization based design and simulator evaluation can be perform simultaneously for each measurement scenario, meaning the computation time can be drastically reduced if sufficient compute resources are available.